\documentclass[10pt, doublecolumn, twoside]{IEEEtran}

\ifCLASSINFOpdf
\else
\fi

\usepackage[T1]{fontenc}
\usepackage{amssymb}
\usepackage{amsmath}
\usepackage{setspace}
\usepackage{algorithm}
\usepackage{algorithmic}
\usepackage{multirow}
\usepackage{url}
\usepackage[numbers,sort&compress]{natbib}

\usepackage{graphicx}
\usepackage{epstopdf}

\usepackage{subfigure}
\usepackage{caption}

\begin{document}
\title{Transmit Beamforming for MISO Broadcast Channels with Statistical and Delayed CSIT}

\author{Mingbo~Dai,~\IEEEmembership{Student Member,~IEEE,}
        Bruno~Clerckx,~\IEEEmembership{Member,~IEEE}
\thanks{M. Dai and B. Clerckx are with the Department of Electrical and Electronic Engineering, Imperial College London, UK, SW7 2AZ UK (e-mail: \{m.dai13, b.clerckx\}@imperial.ac.uk). B. Clerckx is also with the School of Electrical Engineering, Korea University, Seoul, Korea. B. Clerckx is the corresponding author. }}

%
%


\maketitle

\begin{abstract}
This paper focuses on linear beamforming design and power allocation strategy for ergodic rate optimization in a two-user Multiple-Input-Single-Output (MISO) system with statistical and delayed channel state information at the transmitter (CSIT). We propose a transmission strategy, denoted as Statistical Alternative MAT (SAMAT), which exploits both channel statistics and delayed CSIT. Firstly, with statistical CSIT only, we focus on statistical beamforming (SBF) design that maximizes a lower bound on the ergodic sum-rate. Secondly, relying on both statistical and delayed CSIT, an iterative algorithm is proposed to compute the precoding vectors of Alternative MAT (AMAT), originally proposed by Yang et al., which maximizes an approximation of the ergodic sum-rate with equal power allocation. Finally, via proper power allocation, the SAMAT framework is proposed to softly bridge between SBF and AMAT for an arbitrary number of transmit antennas and signal-to-noise ratio (SNR). A necessary condition for the power allocation optimization is identified from the Karush-Kuhn-Tucker (KKT) conditions. The optimum power allocation to maximize an ergodic sum-rate approximation is computed using Sequential Quadratic Programming (SQP). Simulation results show that the proposed SAMAT scheme yields a significant sum-rate enhancement over both SBF and AMAT.

\end{abstract}


%
\IEEEpeerreviewmaketitle

\section{INTRODUCTION}

\IEEEPARstart{I}{n} multi-user (MU) MISO Broadcast Channel (BC), schemes that achieve the sum-rate capacity and the capacity region have been studied in \cite{Joh2005,Wyu2004,Tmichel2006,Jindal2005}. The performance of these optimized approaches heavily depends on the acquisition of accurate and instantaneous Channel State Information (CSI) at the transmitter, which is not feasible in practice due to channel estimation errors, limited feedback resources and delay \cite{Bruno2013}, \cite{maddah2012}. Moreover, obtaining perfect CSIT can incur unaffordable feedback overhead \cite{rajanna2012}.

In practice, statistical CSI and/or perfect but outdated CSI is only available at the transmitter side. The former term which is characterized by the channel covariance matrix varies slowly and thereby can be easily and accurately acquired through long-term feedback. One simple Statistical Beamforming (SBF) approach is to precode the transmitted symbol along the weakest eigenvector of the channel covariance matrix of the unintended user. Such a scheme generalizes the idea of the Perfect CSIT-aided Zero-Forcing Beamforming (PZFBF) to the statistical CSIT only environment. Thus, we denote the SBF strategy with Weakest Eigenvector as SWEBF. Recently, SBF method with Generalized Eigenvector (SGEBF) has been shown to maximize the ergodic sum-rate at high SNR for $M$-user $M$-transmit-antenna MISO BC when $M = 2$ \cite{raghavan2011} or $M \rightarrow \infty$ \cite{choi2013}. However, the optimal precoder for the general $M > 2$ case is still unknown due to a lack of closed-form ergodic sum-rate expression. In \cite{wang2012}, the generalized eigenvector (GE) solution is arrived based on the ergodic signal-to-leakage-and-noise ratio (SLNR), which leverages independence between the numerator and denominator of SLNR. However, the optimal solution to the true problem (ergodic sum-rate) is not apparent. In this paper, we focus on a lower bound on the ergodic sum-rate and address the problem for two-user, arbitrary $M$-transmit-antenna case.

Whenever the feedback delay is larger than the channel's coherence time, the CSIT is delayed but was proved to still benefit the Degree-of-Freedom (DoF) of MISO BC \cite{maddah2012}. In the example of two-transmit-antenna, two-user channel, the maximum sum DoF of $\frac{4}{3}$ can be obtained by retransmitting the overheard interference and doing interference cancellation. This strategy is referred to as MAT. The work \cite{yi2013} generalized the finding of \cite{maddah2012} as GMAT and achieved a higher data rate at finite SNR by constructing precoders which strike a balance between desired signal enhancement and interference alignment. Moreover, an alternative MAT (AMAT) transmission strategy \cite{yang2013} was introduced to achieve a sum DoF between $[\frac{4}{3}, \, 2]$ by utilizing both imperfect current CSIT and perfect delayed CSIT. In time correlated channel, the solution proposed in \cite{yang2013} smoothly bridges between MAT and PZFBF in terms of sum DoF. Inspired by \cite{yang2013}, we aim to bridge between AMAT and SBF in terms of ergodic sum-rate based on statistical and delayed CSIT. Although our framework is similar to \cite{yang2013}, the channel models are completely different. In our scenario, spatially correlated channel model is assumed. Statistical CSIT (full-rank channel covariance matrix) is useless for DoF enhancement or equivalent transmission slot reduction, since the channel estimation error based only on this statistical information does not scale down with SNR. Nevertheless, statistical CSIT is useful for boosting the sum-rate at finite SNR.

With both channel statistics and delayed CSIT at hand, authors in \cite{wang2013} developed an enhanced MAT strategy, denoted as VMAT, yielding a higher sum-rate than the original MAT at finite SNR. However, in highly-correlated channel, the rate performance of VMAT is still inferior to SBF\footnote{Throughout the paper, SBF refers to either SWEBF or SGEBF.} which exploits only statistical CSIT \cite{wang2013}. In a nutshell, statistical channel information is not fully exploited. In \cite{Bruno2014}, authors analyze the error rate performance at finite SNR and the diversity-multiplexing trade-off at infinite SNR of a space-time encoded transmission with delayed and statistical CSIT. So far, there has been no investigation on how to further enhance the finite SNR sum-rate performance beyond that achievable with either statistical CSIT strategies (e.g., SBF) or delayed CSIT strategies (e.g., AMAT). In this paper, we design a spatial precoding AMAT-style transmission scheme, denoted as statistical AMAT (SAMAT), that softly bridges the gap between AMAT and SBF at any SNR. With this background, the main focus of this paper is to investigate power allocation and beamforming optimization. Specifically, the main contributions are listed as follows:

\begin{itemize}
  \item With either statistical or delayed CSIT, the proposed SAMAT strategy can directly boil down to SBF or AMAT. In the former case, we show that SGEBF is optimal to maximize a lower bound on the ergodic sum-rate at high SNR. Under both CSIT but equal power allocation, statistical CSIT-aided AMAT can achieve a significant boost of rate performance relative to the original AMAT. More specifically, an efficient iterative algorithm is developed to compute the optimal statistical precoders to maximize an approximation of the ergodic sum-rate. Monotonic convergence of the algorithm is proved. For two-user two-transmit-antenna case, any two beamforming vectors constituting a unitary matrix are certified to be optimal.

  \item The power allocation can be further optimized to maximize the ergodic sum-rate. A SAMAT transmission strategy is proposed to bridge the gap between SBF and AMAT at any SNR. Due to the complexity of deriving a closed-form expression for the ergodic sum-rate, a tractable approximation needs to be computed. A Sequential Quadratic Programming (SQP) algorithm \cite{bonnans2006} is implemented to solve the consequent non-linear non-convex constrained optimization problem. The necessary condition on power allocation optimization is further identified. In low spatial correlation channels, the proposed SAMAT scheme boils down to AMAT. In highly correlated channels, SAMAT behaves as SBF at low SNR while performing as AMAT at high SNR. In general, SAMAT enables a significantly higher sum-rate than both SBF and AMAT. This achievement stems from two aspects: 1) the transmission of extra private messages; 2) the optimized power allocation for SAMAT.

\end{itemize}

The rest of this paper is organized as follows. Section \ref{systemmodel} introduces the system model. Section \ref{sec:SBF} and \ref{sec:AMAT} elaborate the optimal precoder design for SBF and statistical CSIT-aided AMAT, respectively. In Section \ref{SAMAT_scheme}, we formulate SAMAT and maximize the achievable ergodic sum-rate by optimizing the power allocation. Numerical results are shown in Section \ref{numresults} while Section \ref{conclusion} concludes the paper.

\emph{Notations:} Bold lower case and upper case letters denote vectors and matrices, respectively. The superscripts $(\cdot)^T$ and $(\cdot)^H$ represent the transpose and conjugate transpose. The notation $\text{diag}(\cdot)$ stands for a diagonal matrix whereas $E(\cdot)$ is the expectation operator. $\lambda_{\text{max}}(\cdot)$ and $\lambda_{\text{min}}(\cdot)$ indicate the largest and smallest eigenvalues of a matrix and their corresponding eigenvectors are denoted by $\mathbf{u}_{\text{max}}(\cdot)$ and $\mathbf{u}_{\text{min}}(\cdot)$, respectively. $\mathbb{N}(\cdot)$ stands for the null space of a matrix. Operators $\text{Tr}(\cdot)$ and $\det(\cdot)$ refer to the trace and determinant of a matrix. We denote $\text{Exp}(c)$ as the exponential distribution with parameter $c$ and $U(a,b)$ as the uniform distribution. Let $\Phi_{PD} = \{\mathbf{R} \in \mathbb{C}^{M \times M} \;|\mathbf{R} \text{ is positive definite}\}$.

\section{SYSTEM MODEL}
\label{systemmodel}

\begin{figure*}[t]
\centering
\includegraphics[width = 0.7\textwidth]{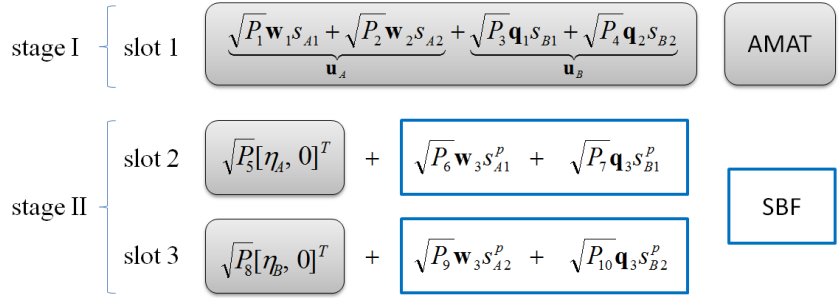}\\
\caption{Block diagram of the proposed SAMAT scheme.} \label{fig:block}
\end{figure*}

Consider a MISO broadcast channel where the transmitter equipped with $M$ antennas ($M \ge$ 2) wishes to send private messages to two users each with a single antenna. Perfect CSI is instantaneously available at the receiver side whereas the transmitter acquires this information with a delay larger than the coherence time of the channel (such that the reported CSI known at the transmitter is uncorrelated with the current CSI). We assume perfect statistical CSIT which is characterized by the spatial correlation matrix. This is a reasonable assumption because channel statistics are more related to the scattering environment and independent of the transmission period.

Rayleigh fading channel model is considered, which implies that the spatial statistics can be completely depicted by the second-order moments of the channel \cite{raghavan2011}. Specifically, we denote the channel between the transmitter and user A in time slot $j$ as $ \mathbf{h}_j$ and similarly $ \mathbf{g}_j$ for user B:

\begin{equation} \label{eq:channel_vector}
  \begin{array}{c}
   \mathbf{h}_j=\mathbf{R}^{1/2}_{\scriptscriptstyle{A}} \, \mathbf{h}_{w,j} \\
   \mathbf{g}_j=\mathbf{R}^{1/2}_{\scriptscriptstyle{B}} \, \mathbf{g}_{w,j},
\end{array}
\end{equation}
where $\mathbf{h}_{w,j}$ and $\mathbf{g}_{w,j}$ are $M \times$1 vectors with independent and identical distribution (i.i.d) $ \mathcal{CN} (0,1)$ entries. They are assumed constant within one time slot and varying independently across time slots. $\mathbf{R}_A$ and $\mathbf{R}_B$ are full rank positive definite covariance matrices\footnote{For rank deficient case, the symbol intended to user $i$ is simply precoded by a column vector in $\mathbb{N}(\mathbf{R}_{\scriptscriptstyle{j}})$. By doing this, the overheard interference of each symbol can be completely removed. Thus, we can transmit two symbols at one time instant, achieving a sum DoF of 2 as if we have perfect CSIT.} for user A and B respectively, which can be decomposed as $\mathbf{R}_k = \mathbf{V}_k \mathbf{\Lambda}_k \mathbf{V}^H_k,\, k = A,B$. $\mathbf{V}_k \in \mathbb{C} ^{M\times M}$ is a unitary matrix whose columns are eigenvectors of $\mathbf{R}_k$, while the diagonal $\mathbf{\Lambda}_k$ that contains the eigenvalues of $\mathbf{R}_k$ is normalized as $\text{Tr}(\mathbf{\Lambda}_k) = M$. $\mathbf{\Lambda}_k = \mathbf{I}$ indicates the $k$-th channel is spatially uncorrelated while $\text{rank}(\mathbf{\Lambda}_k) = 1$ implies it is fully correlated \cite{srinivasan2007}.

The proposed SAMAT framework is shown in Fig. \ref{fig:block}. It contains two stages/three time slots. In the $1^{\text{st}}$ slot, the transmitter superposes four private symbols $s_{\scriptscriptstyle{A1}}, s_{\scriptscriptstyle{A2}}, s_{\scriptscriptstyle{B1}}, s_{\scriptscriptstyle{B2}}$ and sends them to both users. Denote $\mathbf{u}_{\scriptscriptstyle{A}} = \mathbf{W} \mathbf{P}^{1/2}_{\scriptscriptstyle{A}} \mathbf{s}_{\scriptscriptstyle{A}}$ and $\mathbf{u}_{\scriptscriptstyle{B}} = \mathbf{Q} \mathbf{P}^{1/2}_{\scriptscriptstyle{B}} \mathbf{s}_{\scriptscriptstyle{B}}$ as the encoded symbols with statistical beamformer and power allocation, where $\mathbf{W} = [\mathbf{w}_{1} \; \mathbf{w}_{2}], \mathbf{Q} = [\mathbf{q}_{1} \; \mathbf{q}_{2}], \mathbf{P}_{\scriptscriptstyle{A}} = \text{diag}(P_{\scriptscriptstyle{1}}, \, P_{\scriptscriptstyle{2}}), \mathbf{P}_{\scriptscriptstyle{B}} = \text{diag}(P_{\scriptscriptstyle{3}}, \, P_{\scriptscriptstyle{4}})$. $\mathbf{s}_{\scriptscriptstyle{k}} = [s_{\scriptscriptstyle{k1}}, s_{\scriptscriptstyle{k2}}]^T$ represents the Gaussian symbols intended to user $k$ and $E\{\mathbf{s}_{k}\mathbf{s}^H_{k}\} = \mathbf{I}$. At the end of this stage, each user receives its desired signal as well as the overheard interference due to the superposed transmission. Denote $\eta_{\scriptscriptstyle{A}} = \mathbf{h}^H_1 \mathbf{u}_{B}$ and $\eta_{\scriptscriptstyle{B}} = \mathbf{g}^H_1 \mathbf{u}_{A}$ as the interference overheard by user $A$ and $B$, respectively. In stage II, the transmitter has access to $\mathbf{h}_1$ and $\mathbf{g}_1$ (delayed CSIT). Then, $\eta_{\scriptscriptstyle{A}}$ and $\eta_{\scriptscriptstyle{B}}$ can be reconstructed and broadcast via a single antenna in the following two slots. This stage helps both users eliminate the overheard interference and reinforce the desired signals. In addition, new private messages $s^p_{\scriptscriptstyle{A1}}, s^p_{\scriptscriptstyle{A2}}, s^p_{\scriptscriptstyle{B1}}, s^p_{\scriptscriptstyle{B2}}$ are sent to both users in a superposed fashion and this extra transmission makes use of statistical CSIT only. $P_k \ge 0, k = 1,\ldots,10$ indicate the power allocated to each symbol. $\mathbf{w}_{k}$ and $\mathbf{q}_{k}, k = 1,\ldots,3$ denote $M \times$1 unit-norm precoders which depend only on statistical CSIT.

The proposed SAMAT scheme facilitates a smart use of statistical and/or delayed CSIT. With statistical CSIT only, we simply put $P_{\scriptscriptstyle{1}} = P_{\scriptscriptstyle{3}} = P_{\scriptscriptstyle{5}} = P_{\scriptscriptstyle{8}} = 0$ and then SAMAT boils down to SBF in each time slot. With delayed CSIT only, we simply put $P_{\scriptscriptstyle{6}} = P_{\scriptscriptstyle{7}} = P_{\scriptscriptstyle{9}} = P_{\scriptscriptstyle{10}} = 0$. Then, SAMAT becomes the AMAT scheme and enables a sum DoF of $\frac{4}{3}$ at high SNR. If the transmitter has both statistical and delayed CSIT, proper power allocation and statistical precoding can make room for extra symbols transmission. The benefits of transmitting new symbols overcome the loss caused by the interference it creates to the main AMAT transmission. In this case, the proposed SAMAT framework allows for the parallel transmission of SBF on top of AMAT while outperforming AMAT and SBF at any SNR. More specifically, the transmitted signals are written as

\vspace{-10pt}

\begin{equation} \label{eq:tx_signal}
\begin{array}{lcl}
\mathbf{x}_1 &=& \mathbf{u}_{\scriptscriptstyle{A}} + \mathbf{u}_{\scriptscriptstyle{B}}\\
\mathbf{x}_2 &=&  \sqrt{P_{\scriptscriptstyle{5}}} \begin{bmatrix} \eta_{\scriptscriptstyle{A}}, \; 0 \end{bmatrix}^T \; \hspace{-1pt} \hphantom{~}+\hphantom{~} \sqrt{P_{\scriptscriptstyle{6}}} \, \mathbf{w}_3 \, s^p_{\scriptscriptstyle{A1}} \hspace{2pt} \hphantom{~}+\hphantom{~} \sqrt{P_{\scriptscriptstyle{7}}} \, \mathbf{q}_3 \, s^p_{\scriptscriptstyle{B1}} \\
\mathbf{x}_3 &=&  \sqrt{P_{\scriptscriptstyle{8}}} \begin{bmatrix} \eta_{\scriptscriptstyle{B}}, \; 0 \end{bmatrix}^T \hspace{2pt} \hphantom{~}+\hphantom{~} \sqrt{P_{\scriptscriptstyle{9}}} \, \mathbf{w}_3 \, s^p_{\scriptscriptstyle{A2}} \hphantom{~}+\hphantom{~}  \sqrt{P_{\scriptscriptstyle{10}}} \, \mathbf{q}_3 \, s^p_{\scriptscriptstyle{B2}}.
\end{array}
\end{equation}

For (A)MAT-based schemes, the transmit power in stage II is inherently dependent on the channel realization that changes rapidly. The power consumption in each transmission period hardly keeps constant. Thus, a less restrictive metric is the long-term average power constraint

\vspace{-8pt}

\begin{eqnarray} \label{eq:power_const}
\bar{P}_{\text{c}} &=& E[\text{Tr}(\mathbf{x}_1 \mathbf{x}^H_1)] + E[\text{Tr}(\mathbf{x}_2 \mathbf{x}^H_2)] + E[\text{Tr}(\mathbf{x}_3 \mathbf{x}^H_3)] \nonumber \\ \nonumber
&=& P_{\scriptscriptstyle{1}}+P_{\scriptscriptstyle{2}}+P_{\scriptscriptstyle{3}}+P_{\scriptscriptstyle{4}}+ P_{\scriptscriptstyle{6}} + P_{\scriptscriptstyle{7}} + P_{\scriptscriptstyle{9}} + P_{\scriptscriptstyle{10}} + \\ \nonumber && P_{\scriptscriptstyle{5}}(\lambda_{\scriptscriptstyle{A1}}P_{\scriptscriptstyle{3}} + \lambda_{\scriptscriptstyle{A2}} P_{\scriptscriptstyle{4}}) + P_{\scriptscriptstyle{8}}(\lambda_{\scriptscriptstyle{B1}}P_{\scriptscriptstyle{1}} + \lambda_{\scriptscriptstyle{B2}} P_{\scriptscriptstyle{2}})\\
&\le& 3P,
\end{eqnarray}
where $\lambda_{\scriptscriptstyle{A1}} = \mathbf{q}^H_1 \mathbf{R}_{\scriptscriptstyle{A}} \mathbf{q}_1, \, \lambda_{\scriptscriptstyle{A2}} = \mathbf{q}^H_2 \mathbf{R}_{\scriptscriptstyle{A}} \mathbf{q}_2, \, \lambda_{\scriptscriptstyle{B1}} = \mathbf{w}^H_1 \mathbf{R}_{\scriptscriptstyle{B}} \mathbf{w}_1, \, \lambda_{\scriptscriptstyle{B2}} = \mathbf{w}^H_2 \mathbf{R}_{\scriptscriptstyle{B}} \mathbf{w}_2$.
The expectation is taken over the input signals and the channels and $P$ denotes the average power budget of the transmitter for each time slot.

Although our framework is similar to \cite{yang2013}, there are essential distinctions between them. A first distinction lies in the channel model. We exploit spatial correlation to compress the interference and make room for extra symbols transmission while they make use of time correlation. The power allocation in \cite{yang2013} depends on SNR and quality of current CSIT while our power allocation strategy relies on SNR, precoder design as well as spatial correlation.

The secondary distinction lies in the encoding/decoding strategy (and hence the transmission protocol). More specifically, interference quantization is crucial for \cite{yang2013}, where the overheard interference symbol with a reduced power is transmitted with full power in order to save channel resources. Interference quantization is proposed to solve the consequent problem of power mismatch (which scales with transmit power). By decoding the interference symbols first, \cite{yang2013} equivalently obtains one AMAT transmission plus two ZF transmissions. DoF gain at high SNR can be obtained over the original AMAT scheme.

In contrast with \cite{yang2013}, the overheard interference is multicast by analog transmission in our scenario and the reason is threefold. First, we retransmit the interference symbols after scaling them by constant (i.e., not scaling with the transmit power) factors $P_{\scriptscriptstyle{5}}$ and $P_{\scriptscriptstyle{8}}$. $P_{\scriptscriptstyle{1,2,3,4}}$ in the main AMAT transmission scale with the transmit power at high SNR to achieve the DoF of $\frac{4}{3}$. However, to guarantee the power constraint, the multiplication terms $P_{\scriptscriptstyle{5}}(\lambda_{\scriptscriptstyle{A1}}P_{\scriptscriptstyle{3}} + \lambda_{\scriptscriptstyle{A2}} P_{\scriptscriptstyle{4}})$ and $P_{\scriptscriptstyle{8}}(\lambda_{\scriptscriptstyle{B1}}P_{\scriptscriptstyle{1}} + \lambda_{\scriptscriptstyle{B2}} P_{\scriptscriptstyle{2}})$ in eq. \eqref{eq:power_const} limit $P_{\scriptscriptstyle{5}}$ and $P_{\scriptscriptstyle{8}}$ to some constants. Second, interference quantization would prevent the proposed SAMAT scheme from bridging SBF at low SNR. More specifically, in the $2^{\text{nd}}/3^{\text{rd}}$ slot, SAMAT should behave as SBF at low SNR and should therefore allocate most of the transmit power to the extra symbols and only a little power to the overheard interference symbol. Following \cite{yang2013}, if interference quantization is applied, the digitized interference should be decoded first by treating the extra symbols as noise. In this case, however, the decoding would fail because the noise power would overwhelm the desired signal power. Third, due to the inherent properties of the channel model (full-rank channel covariance matrix), a sum DoF strictly larger than $\frac{4}{3}$ cannot be achieved in our case (contrary to \cite{yang2013}). Hence the SAMAT transmission and reception strategies are not motivated by a DoF maximization. With SAMAT, a sum DoF of $\frac{4}{3}$ is achieved where the extra private symbols are not used to increase the DoF at high SNR (contrary to \cite{yang2013}) but to boost the sum-rate at low/finite SNR. This implies that the retransmitted overheard interference does not have to be decoded first in SAMAT (contrary to \cite{yang2013}) but can simply be aligned and cancelled so as to decode the private and extra symbols. Recall again that \cite{yang2013} relies on interference quantization to decode first the overheard interference and then the private messages in order to increase the sum DoF beyond $\frac{4}{3}$. The detailed description of our decoding strategy is provided in section \ref{SAMAT_scheme}.

\section{STATISTICAL BEAMFORMING}
\label{sec:SBF}

Let us first focus on the scenario with statistical CSIT only. As mentioned before, the proposed SAMAT scheme boils down to SBF, i.e., the transmitter sends two statistically precoded symbols, each intended to one user. Since the system model for the transmission in each slot is identical, we can focus on one slot and omit the subscript. Assume the transmitter equally allocates its power budget to both users, the delivered signal described by \eqref{eq:tx_signal} becomes
\begin{equation} \label{eq:szfbf}
\mathbf{x} = \sqrt{\rho}  \, \mathbf{w} \, s_{\scriptscriptstyle{A}} \hphantom{~}+\hphantom{~} \sqrt{\rho}  \, \mathbf{q} \, s_{\scriptscriptstyle{B}},
\end{equation}
where $\rho=\frac{P}{2}$. For simplicity, we will look at the rate performance of user A only and a similar derivation can be easily extended to user B. The received signal at the receiver side is given as $y=\sqrt{\rho}\,\mathbf{h}^H \mathbf{w} \,s_{\scriptscriptstyle{A}} + \sqrt{\rho} \,\mathbf{h}^H \mathbf{q} \,s_{\scriptscriptstyle{B}} + n_{\scriptscriptstyle{A}}$, where $\mathbf{h}\in \mathbb{C}^M$ is the channel vector and $n_{\scriptscriptstyle{A}} \sim \mathcal{CN} (0,1)$ is the standard complex additive white Gaussian noise (AWGN). The achievable ergodic rate of user A is given by

\begin{equation} \label{eq:ergodic_rate}
R_{\scriptscriptstyle{A}} = E \left[\log_2 \left(1+\text{SINR}_{\scriptscriptstyle{A}}\right)\right],
\end{equation}
where $\text{SINR}_{\scriptscriptstyle{A}} = \frac{\rho | \mathbf{h}^H \mathbf{w} |^2}{1 +  \rho | \mathbf{h}^H \mathbf{q} |^2}$ is the instantaneous signal-to-interference-and-noise ratio (SINR) by treating the multi-user interference as noise. Consequently, the ergodic sum-rate of the system with linear beamforming is $R_{\scriptstyle{\text{sum}}} \triangleq R_{\scriptscriptstyle{A}} + R_{\scriptscriptstyle{B}}$.

In the spatially correlated channels, the role of spatial statistics on the rate performance of a linear beamformer has been well studied in \cite{raghavan2011}. They consider a MISO broadcast setting with $M$ antennas at the transmitter and $M$ single-antenna users. When $M =$ 2, optimal precoders that maximize the ergodic sum-rate were developed at extremely high/low SNR. Meanwhile, the best precoding vectors are still unknown for general case ($M >$ 2) due to the difficulty of computing a closed-form expression for the ergodic sum-rate. In this paper, we show by the following theorem that the generalized eigenvector (GE) is still valid for the two-user, $M$-transmit-antenna scenario.

$\textbf{Theorem 1:}$ For any $M$, the ergodic sum-rate of SBF at high SNR can be lower bounded by

\begin{equation} \label{eq:SBF_LB}
R_{\scriptstyle{\text{sum}}} \ge \log_2 \left( \frac{\mathbf{w}^H \mathbf{R}_{\scriptscriptstyle{A}} \mathbf{w} }{\mathbf{w}^H \mathbf{R}_{\scriptscriptstyle{B}} \mathbf{w} } \frac{\mathbf{q}^H \mathbf{R}_{\scriptscriptstyle{B}} \mathbf{q} }{\mathbf{q}^H \mathbf{R}_{\scriptscriptstyle{A}} \mathbf{q} }\right),
\end{equation}
where $\gamma$ is the Euler constant. The precoders that maximize the lower bound in \eqref{eq:SBF_LB} are given by
\vspace{-8pt}

\begin{equation} \label{eq:GE_precoder}
\mathbf{w}_{\scriptscriptstyle{GE}} = \mathbf{u}_\text{max}(\mathbf{R}^{-1}_{\scriptscriptstyle{B}} \mathbf{R}_{\scriptscriptstyle{A}}), \quad
\mathbf{q}_{\scriptscriptstyle{GE}} = \mathbf{u}_\text{max}(\mathbf{R}^{-1}_{\scriptscriptstyle{A}} \mathbf{R}_{\scriptscriptstyle{B}}).
\end{equation}

The corresponding lower bound of the ergodic sum-rate is

\begin{equation} \label{eq:lbrate}
\begin{array}{lcl}
R_{\text{sum,lb}} &=& \log_2\left({\chi{\left(\mathbf{R}^{-1}_{\scriptscriptstyle{B}}\mathbf{R}_{\scriptscriptstyle{A}}\right)}}\right) = \log_2\left({\chi{\left(\mathbf{R}^{-1}_{\scriptscriptstyle{A}}\mathbf{R}_{\scriptscriptstyle{B}}\right)}}\right),
\end{array}
\end{equation}
where $\chi(\mathbf{\cdot}) = \frac{\lambda_{\text{max}}(\mathbf{\cdot})}{\lambda_{\text{min}}(\mathbf{\cdot})}$ is the condition number.

\vspace{3pt}
\begin{IEEEproof}
A detailed proof is relegated in Appendix \ref{sec:proof_th1}.
\end{IEEEproof} \vspace{1pt}
A special case\footnote{When the two users share the same set of statistical eigenmodes but orthogonal dominant eigenvectors and $M =$ 2.} of \eqref{eq:lbrate} is confirmed by \cite[Corollary 2]{wang2012} by noting that $\chi(\mathbf{\cdot}) > 1$. In the low SNR region where the interference can be completely ignored, the optimal choice is to send along the dominant statistical eigen-mode of the user's own channel \cite{veeravalli2005}. At intermediate SNR, however, \cite{raghavan2011} has shown the difficulty of finding a closed-form expression of the optimal precoders even for $M =$ 2 case. Instead, it is solved by an exhaustive search operated upon a linearly combined high- and low-SNR solution. In the general $M >$ 2 case, we compute only a high-SNR solution and avoid the line search method. The simulation results show that it works well at practical SNR.

\emph{Remark 1:} The closed-form precoders that maximize the ergodic sum-rate of SBF is difficult to compute due to the coupled nature in SINR expression. To solve this problem, we can use an alternative SLNR metric, which is defined as $\text{SLNR}_{\scriptscriptstyle{A}} = \frac{\rho | \mathbf{h}^H \mathbf{w} |^2}{1 +  \rho | \mathbf{g}^H \mathbf{w} |^2}$. At high SNR, the maximization of a lower bound on $E[\text{SLNR}_{\scriptscriptstyle{A}}]$ also leads to the solution \eqref{eq:GE_precoder}. Similarly, the effectiveness of the SLNR metric in designing multi-user transmit beamforming vectors has been examined in  \cite{sadek2007,wang2012,raghavan2011}.

In contrast with SGEBF, the precoding vectors of SWEBF in correlated channel can be written as $\mathbf{w}_{\scriptscriptstyle{WE}} = \mathbf{u}_{\text{min}}(\mathbf{R}_{\scriptscriptstyle{B}} ), \, \mathbf{q}_{\scriptscriptstyle{WE}} = \mathbf{u}_{\text{min}}(\mathbf{R}_{\scriptscriptstyle{A}} )$. However, the rate performance of SWEBF is unfavourable in the scenario where both channels of user A and B have similar weakest eigen-direction (e.g., co-located users). To be specific, the precoding vectors which are designed to remove the interference also cancel out the intended signal. By contrast, the GE beamforming approach obtains a balance between interference cancellation and desired signal enhancement. In other words, SGEBF exhibits robustness with respect to different channels compared to SWEBF. Moreover, we can observe from the numerical results that SGEBF outperforms SWEBF further as $M$ increases.

\section{ALTERNATIVE MAT}
\label{sec:AMAT}

Let us focus on the scenario with both delayed and statistical CSIT. Under equal power allocation, \cite{wang2013} has shown that additional channel statistics enable a higher achievable sum-rate relative to the original MAT. However, there are big differences between this section and \cite{wang2013}. First, they released the power constraint in the interference retransmission phase (stage II), which leads to a variation of the total transmit power. We here control the power consumption by using a long-term power constraint. Second, an efficient iterative algorithm is developed to compute the statistical precoders to maximize an approximation of the ergodic sum-rate. Particularly, monotonic convergence of the algorithm is proved. This section reveals how much the ergodic sum-rate can be improved by statistical precoder only. It will be used as a baseline for the next section where the rate performance is further enhanced by additional power allocation.

\subsection{Rate Approximation}
The proposed SAMAT scheme becomes statistical CSIT-aided AMAT by simply letting $P_{\scriptscriptstyle{6}} = P_{\scriptscriptstyle{7}} = P_{\scriptscriptstyle{9}} = P_{\scriptscriptstyle{10}} = 0, P_{\scriptscriptstyle{5}} = P_{\scriptscriptstyle{8}} = 1$ in \eqref{eq:tx_signal} and the transmitter distributes equal power to four symbols, denoted as $\rho$. The transmitted signals can be expressed as

\vspace{-2pt}

\begin{equation} \label{eq:AMAT_signal}
\begin{array}{lcl}
\mathbf{x}_1 &=& \sqrt{\rho} \, \mathbf{W} \, \mathbf{s}_{\scriptscriptstyle{A}} \hphantom{~}+\hphantom{~}
                 \sqrt{\rho} \, \mathbf{Q} \, \mathbf{s}_{\scriptscriptstyle{B}} \\
\mathbf{x}_2 &=& \sqrt{\rho} \begin{bmatrix} \mathbf{h}^H_1\mathbf{Q} \, \mathbf{s}_{\scriptscriptstyle{B}}, \; 0 \end{bmatrix}^T \; \\
\mathbf{x}_3 &=& \sqrt{\rho} \begin{bmatrix} \mathbf{g}^H_1\mathbf{W} \,  \mathbf{s}_{\scriptscriptstyle{A}}, \; 0 \end{bmatrix}^T.
\end{array}
\end{equation}

With delayed CSIT only, $\mathbf{W} = \mathbf{Q} = \mathbf{I}$ and the proposed SAMAT scheme further boils down to the original AMAT scheme. For simplicity, we focus on the performance of user A and similar results can be symmetrically applied to user B. The signal vector received by user A is given by

\vspace{-2pt}

\begin{equation} \label{eq:amat}
\mathbf{y}_{\scriptscriptstyle{A}} = \sqrt{\rho} \begin{bmatrix} \mathbf{h}^H_1 \mathbf{W} \\ \mathbf{0} \\ h^{\ast}_{\scriptscriptstyle{31}} \mathbf{g}^H_1 \mathbf{W} \end{bmatrix}\mathbf{s}_{\scriptscriptstyle{A}} +
\sqrt{\rho} \begin{bmatrix} \mathbf{h}^H_1 \mathbf{Q} \\ h^{\ast}_{\scriptscriptstyle{21}} \mathbf{h}^H_1 \mathbf{Q}  \\ \mathbf{0} \end{bmatrix}\mathbf{s}_{\scriptscriptstyle{B}} +
\begin{bmatrix} n_{\scriptscriptstyle{A1}} \\ n_{\scriptscriptstyle{A2}} \\ n_{\scriptscriptstyle{A3}} \end{bmatrix},
\end{equation}
where $\mathbf{y}_{\scriptscriptstyle{A}} \triangleq [y_{\scriptscriptstyle{A1}},y_{\scriptscriptstyle{A2}},y_{\scriptscriptstyle{A3}}]^T$ denotes the received signals over three time slots and $h_{jm}$ denotes the channel coefficient between $m$-th transmit antenna and user A in time slot $j$. $n_{\scriptscriptstyle{A}j} \sim \mathcal{CN} (0,1)$ is the normalized complex AWGN. After further interference elimination

\begin{equation} \label{eq:amat_tx_signal}
\widetilde{\mathbf{y}}_{\scriptscriptstyle{A}} = \sqrt{\rho} \, \widetilde{\mathbf{H}} \, \mathbf{s}_{\scriptscriptstyle{A}} + \begin{bmatrix}  h^{\ast}_{\scriptscriptstyle{21}} n_{\scriptscriptstyle{A1}} - n_{\scriptscriptstyle{A2}} \\ n_{\scriptscriptstyle{A3}}  \end{bmatrix},
\end{equation}
where $\widetilde{\mathbf{H}} = [(h^{\ast}_{\scriptscriptstyle{21}} \mathbf{h}^H_1 \mathbf{W})^T, \, (h^{\ast}_{\scriptscriptstyle{31}} \mathbf{g}^H_1 \mathbf{W})^T]^T$. $h_{jm}$ denotes the channel coefficient between $m$-th transmit antenna and user A in time slot $j$. By using a Minimum Mean-Square Error (MMSE) receiver with Successive Interference Cancellation (SIC), the ergodic rate achieved per slot by user A is written as:

\vspace{-2pt}

\begin{equation} \label{eq:amat_rate}
R_{\scriptscriptstyle{A}} = \frac{1}{3} \, E \left[ \log_2  \det\left( \mathbf{I} + \rho \widetilde{\mathbf{H}}^H \mathbf{K}^{-1} \widetilde{\mathbf{H}} \right)\right],
\end{equation}
where $\mathbf{K}$ is the covariance matrix of the noise vector in \eqref{eq:amat_tx_signal} and given by $\text{diag}(1 + |h_{\scriptscriptstyle{21}}|^{2},1)$. It is challenging to obtain the closed-form expression of the ergodic rate, especially for $M >$ 2 case. Hence, we optimize the linear beamforming vectors based on an analytical approximation of $R_{\scriptscriptstyle{A}}$, which is given by the following proposition.

$\textbf{Proposition 1:}$ In spatially correlated Rayleigh fading channel, the ergodic rate of user A for AMAT can be approximated as

\begin{equation} \label{eq:amat_rate_appr}
R_{\scriptscriptstyle{A}} \approx \frac{2}{3} \log_2 \left(1 + \rho \sqrt{e^a \Theta}_{\scriptscriptstyle{A}} \right),
\end{equation}
where \vspace{-5pt}
\begin{eqnarray} \label{eq:theta}
\Theta_{\scriptscriptstyle{A}} &=& \text{Tr}(\mathbf{W}^H \mathbf{R}_{\scriptscriptstyle{A}} \mathbf{W}) \text{Tr}(\mathbf{W}^H \mathbf{R}_{\scriptscriptstyle{B}} \mathbf{W}) - \nonumber\\
&& \text{Tr}(\mathbf{W}^H \mathbf{R}_{\scriptscriptstyle{A}} \mathbf{W} \mathbf{W}^H \mathbf{R}_{\scriptscriptstyle{B}} \mathbf{W})
\end{eqnarray} and $a = e \, \text{Ei}(-1) - 2\gamma$, $\text{Ei}(x) = -\int^{\infty}_{-x} \frac{e^{-t}}{t}dt$ is the exponential integral.

\begin{IEEEproof}
See Appendix \ref{sec:proof_prop1}.
\end{IEEEproof}

Then, we can obtain that $R_{\text{sum}} = R_{\scriptscriptstyle{A}} + R_{\scriptscriptstyle{B}} \approx \frac{2}{3} \log_2 \left(1 + \rho \sqrt{e^a \Theta}_{\scriptscriptstyle{A}} \right) + \frac{2}{3} \log_2 \left(1 + \rho \sqrt{e^a \Theta}_{\scriptscriptstyle{B}} \right) $, where $\Theta_{\scriptscriptstyle{B}} = \text{Tr}(\mathbf{Q}^H \mathbf{R}_{\scriptscriptstyle{A}} \mathbf{Q}) \text{Tr}(\mathbf{Q}^H \mathbf{R}_{\scriptscriptstyle{B}} \mathbf{Q}) - \text{Tr}(\mathbf{Q}^H \mathbf{R}_{\scriptscriptstyle{A}} \mathbf{Q} \mathbf{Q}^H \mathbf{R}_{\scriptscriptstyle{B}} \mathbf{Q})$. It shows that the AMAT scheme exploiting delayed CSIT enables a DoF of $\frac{4}{3}$ at high SNR, while the beamforming based on statistical CSIT makes no contribution to the DoF gain. However, the ergodic rate performance at practical SNR benefits from such spatial correlation information. Observe in \eqref{eq:amat_rate_appr} and \eqref{eq:theta} that the ergodic rate relies on the precoders and the spatial correlation matrices. The latter terms are invariable in a certain environment. Therefore, $R_{\scriptscriptstyle{A}}$ and $R_{\scriptscriptstyle{B}}$ only depend on $\Theta_{\scriptscriptstyle{A}}$ ($\mathbf{W}$) and $\Theta_{\scriptscriptstyle{B}}$ ($\mathbf{Q}$), respectively. To maximize the ergodic sum-rate performance, the precoders $\mathbf{W}$ and $\mathbf{Q}$ can be independently designed. Let us focus on $\Theta_{\scriptscriptstyle{A}}$ only and optimize $\mathbf{W}$. Similarly, we can obtain the optimal $\mathbf{Q}$ that maximizes $\Theta_{\scriptscriptstyle{B}}$.

\subsection{Precoder Design}

\subsubsection{Multi-antenna case (M > 2)}

\begin{table}[t]
\caption{Algorithm 1\&2: Precoder Optimization for AMAT}
\label{Iterative_alg}
\begin{tabular}{l}
\hline
1: \textbf{Initialize}: Set iteration index $m = 0$, randomly generate \\
\qquad \qquad \quad \;    and then normalize $\mathbf{w}^{(0)}_1$, $\mathbf{w}^{(0)}_2$\\
2: \textbf{Repeat}\\
3: $m \gets m+1$ \\
4: \quad Update $\mathbf{w}^{(m)}_1$ with GradAct \textbf{[Algorithm 1]}, or\\
\qquad \qquad \qquad \qquad  with Max-Eig \textbf{[Algorithm 2]} \\
5: \quad Update $\mathbf{w}^{(m)}_2$ with GradAct or with Max-Eig \\
6: \textbf{Until} $|\Theta^{(m)}_{\scriptscriptstyle{A}} - \Theta^{(m-1)}_{\scriptscriptstyle{A}}| \le \epsilon$ \\
\hline
\end{tabular}
\end{table}

It is difficult to obtain a closed-form expression of the beamforming vectors that maximize \eqref{eq:theta} and further \eqref{eq:amat_rate_appr}. For such a problem where joint optimization is difficult but the objective function is convex in each of the optimization variables $\mathbf{w}_1$ and $\mathbf{w}_2$, an alternating algorithm, also known as Block Coordinate Descent method, has been widely used in optimization \cite{Xu2013,Tseng2009}. More specifically, we maximize \eqref{eq:theta} by sequentially fixing one vector and updating the other. Fix $\mathbf{w}_2$ and focus on $\mathbf{w}_1$ (vice versa, the following derivations still hold). We can reformulate the subproblem as

\begin{equation} \label{eq:theta2}
\begin{array}{lcl}
\text{max} \, \Theta_{\scriptscriptstyle{A}}(\mathbf{w}_1) = \mathbf{w}_1^H \mathbf{R}_{\scriptscriptstyle{A}} \mathbf{w}_1 \mathbf{w}_2^H \mathbf{R}_{\scriptscriptstyle{B}} \mathbf{w}_2 + \mathbf{w}_1^H \mathbf{R}_{\scriptscriptstyle{B}} \mathbf{w}_1 \mathbf{w}_2^H \mathbf{R}_{\scriptscriptstyle{A}} \mathbf{w}_2  \\
\qquad \qquad \quad \;\; - \; \mathbf{w}_1^H \mathbf{R}_{\scriptscriptstyle{A}} \mathbf{w}_2 \mathbf{w}_2^H \mathbf{R}_{\scriptscriptstyle{B}} \mathbf{w}_1 - \mathbf{w}_1^H \mathbf{R}_{\scriptscriptstyle{B}} \mathbf{w}_2 \mathbf{w}_2^H \mathbf{R}_{\scriptscriptstyle{A}} \mathbf{w}_1  \\
\text{s.t.} \quad \|\mathbf{w}_1\| = 1.
\end{array}
\end{equation}

Since it is convex in $\mathbf{w}_1$\footnote{The convexity can be easily proved with the second order condition, which is omitted here for conciseness.}, the classical gradient ascent (GradAct) method can be used to determine the optimal solution (step 4 of Table I). Once the optimal $\mathbf{w}_1$ is obtained in terms of certain $\mathbf{w}_2$, the process is repeated the other way around (step 5), leading to an iterative algorithm. Since the steepest ascent direction acts as the best direction to increase the objective function, a proper step size can be computed for a non-decreasing objective value, i.e., $\Theta^{(m,4)}_{\scriptscriptstyle{A}} \le \Theta^{(m,5)}_{\scriptscriptstyle{A}} \le \Theta^{(m+1,4)}_{\scriptscriptstyle{A}}$ where $\Theta^{(m,4)}_{\scriptscriptstyle{A}}$ refers to the objective value at step 4 in the $m$-th iteration in Table \ref{Iterative_alg} (Algorithm 1). Thus, the convergence of Algorithm 1 is ensured, since $\Theta_{\scriptscriptstyle{A}}$ is monotonically increased (non-decreased) after each iteration and upper bounded. Even though the optimal solution is obtained for each subproblem, the iterative algorithm can not guarantee the global optimal beamforming vectors.

\begin{figure}[t]
  \centering
  \includegraphics[width = 0.5\textwidth]{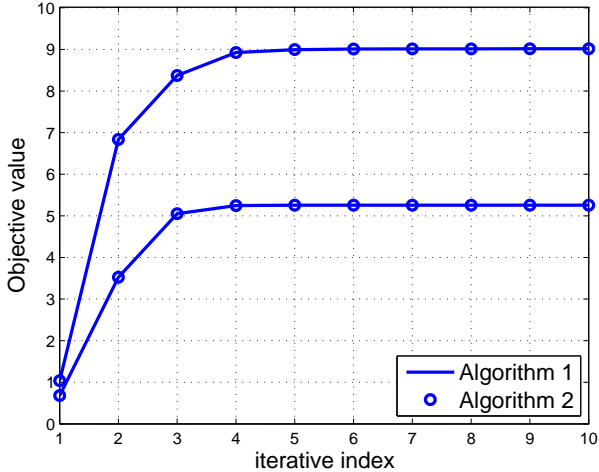}\\
  \caption{Convergence of iterative algorithm 1$\&$2.} \label{fig:converge}
\end{figure}

Alternatively, \eqref{eq:theta2} is quadratic in $\mathbf{w}_1$ and the optimal solution can be obtained by eigen-decomposition. Rewrite \eqref{eq:theta2} as $\Theta_{\scriptscriptstyle{A}}(\mathbf{w}_1) = \mathbf{w}_1^H \, \mathbf{M}(\mathbf{w}_2) \, \mathbf{w}_1$, where $\mathbf{M}(\mathbf{w}_2) = \mathbf{w}_2^H \mathbf{R}_{\scriptscriptstyle{B}} \mathbf{w}_2 \mathbf{R}_{\scriptscriptstyle{A}} + \mathbf{w}_2^H \mathbf{R}_{\scriptscriptstyle{A}} \mathbf{w}_2 \mathbf{R}_{\scriptscriptstyle{B}} - \mathbf{R}_{\scriptscriptstyle{A}} \mathbf{w}_2 \mathbf{w}_2^H \mathbf{R}_{\scriptscriptstyle{B}}$ $- \mathbf{R}_{\scriptscriptstyle{B}} \mathbf{w}_2 \mathbf{w}_2^H \mathbf{R}_{\scriptscriptstyle{A}}$. The closed-form solution is the maximum eigenvector (Max-Eig),

\begin{equation} \label{eq:theta4}
\mathbf{w}_1 = \underset{\|\mathbf{w}_1\|=1}{\text{arg max}} \; \Theta_{\scriptscriptstyle{A}}(\mathbf{w}_1) = \mathbf{u}_{\text{max}} \left(\mathbf{M \left(\mathbf{w}_2 \right)} \right).
\end{equation}

With \eqref{eq:theta4} at hand, we can easily compute the optimal precoders by the proposed iterative approach. In Fig.\ref{fig:converge}, we show by two cases ($M = 4,8$) that the iterative algorithm converges very fast, where the covariance matrices are randomly generated.

\subsubsection{Two-antenna case (M = 2)}
A special case of considerable interest is the two transmit antenna scenario. The optimal precoders can be easily obtained by the following proposition.

$\textbf{Proposition 2:}$ For $M = 2$ MISO BC with spatially correlated Rayleigh fading, any unitary beamforming matrix is optimal to maximize \eqref{eq:theta} and further \eqref{eq:amat_rate_appr}.

\begin{IEEEproof} For arbitrary $M$ and $\mathbf{w}_2$, it is easy to verify that $\mathbf{w}^H_2 \, \mathbf{M}(\mathbf{w}_2) \, \mathbf{w}_1 = 0$, i.e., $\mathbf{M}(\mathbf{w}_2) \, \mathbf{w}_1 \in \mathbb{N}(\mathbf{w}_2)$. The maximization of $\Theta_{\scriptscriptstyle{A}}(\mathbf{w}_1) = \mathbf{w}_1^H \, \mathbf{M}(\mathbf{w}_2) \, \mathbf{w}_1$ leads to the observation that the optimum $\mathbf{w}_1 \in \mathbb{N}(\mathbf{w}_2)$. Similarly, when we fix $\mathbf{w}_1$ and update $\mathbf{w}_2$, we have the optimum $\mathbf{w}_2 \in \mathbb{N}(\mathbf{w}_1)$. It implies that the optimal beamforming vectors are always orthogonally chosen ($\mathbf{w}_1 \perp \mathbf{w}_2$). For the special $M = 2$ case, since $\mathbf{w}_1$ is uniquely defined in $\mathbb{N}(\mathbf{w}_2)$ and vice versa, any two beamforming vectors constituting a unitary matrix are optimal. Moreover, eq. \eqref{eq:theta} becomes constant $\Theta_{\scriptscriptstyle{A}} = \text{Tr}(\mathbf{R}_{\scriptscriptstyle{A}}) \text{Tr}(\mathbf{R}_{\scriptscriptstyle{B}}) - \text{Tr} (\mathbf{R}_{\scriptscriptstyle{A}} \mathbf{R}_{\scriptscriptstyle{B}})$.
\end{IEEEproof}

This proposition reveals that any orthogonal beamforming vectors with equal power allocation achieve the same ergodic sum-rate performance, which is verified by the Fig. \ref{fig_AMATco2} in section \ref{SAMAT_scheme}. Then, let us compute the equal power allocation $\rho$. According to \eqref{eq:power_const}, the long-term average power consumption for AMAT is represented by

\vspace{-5pt}

\begin{flalign} 
\bar{P}_{\text{c}} =& 4\rho + \rho \, \text{Tr}(\mathbf{Q}^H \mathbf{R}_{\scriptscriptstyle{B}} \mathbf{Q} ) + \rho \, \text{Tr}(\mathbf{W}^H \mathbf{R}_{\scriptscriptstyle{A}} \mathbf{W} ) \nonumber\\
                   =& \rho \, (4 + \mathbf{q}_1^H \mathbf{R}_{\scriptscriptstyle{B}} \mathbf{q}_1 + \mathbf{q}_2^H \mathbf{R}_{\scriptscriptstyle{B}} \mathbf{q}_2 + \mathbf{w}_1^H \mathbf{R}_{\scriptscriptstyle{A}} \mathbf{w}_1 + \mathbf{w}_2^H \mathbf{R}_{\scriptscriptstyle{A}} \mathbf{w}_2) \nonumber \\
                   \le& \rho \, \left( 4 + 2M\right). \label{power_cost1}
\end{flalign}
where the inequality (\ref{power_cost1}) is obtained by using $\mathbf{x}_1^H \mathbf{R} \mathbf{x}_1 + \mathbf{x}_2^H \mathbf{R} \mathbf{x}_2 \le \lambda_1(\mathbf{R}) + \lambda_2(\mathbf{R})  \le \text{Tr} \left(\mathbf{R}\right)$, where unit-norm $\mathbf{x}_i$ are mutually orthogonal and $\lambda_i(\mathbf{R})$ corresponds to the $i$-th largest eigenvalue \cite{coope2000}, \cite{horn2012}. In order to maintain the power constraint, equal power allocation is calculated as $\rho = \frac{3P}{4 + 2M}$ (e.g., $\rho = \frac{3P}{8}$ for two transmit antennas). Equality in \eqref{power_cost1} holds for $M = 2$ case, which also justifies proposition 2 in the sense that orthonormal precoders (optimally) use up all the power budget.

\section{STATISTICAL AMAT}
\label{SAMAT_scheme}

In the last section, we explored how the statistical CSIT aids in AMAT under equal power allocation. The corresponding rate performance is superior to the original AMAT, however, still inferior to SBF in highly correlated channels at low/intermediate SNR regime. In fact, the ergodic sum-rate can be further enhanced by power allocation optimization. The proposed SAMAT scheme in Fig. \ref{fig:block} is developed to overcome both SBF and AMAT at any SNR in arbitrary spatial condition. Proper power allocation and statistical precoder design can compress the received interference $\eta_{\scriptscriptstyle{A}} (\eta_{\scriptscriptstyle{B}})$ in $1^{\text{st}}$ slot (as well as common symbols in $2^{\text{nd}}/3^{\text{rd}}$ slot). Meanwhile, it makes room for new symbol transmission in stage II which brings more rate benefits. To achieve this, the power allocation is optimized based on a tractable approximation of the ergodic sum-rate of SAMAT. The power allocation strategy that maximizes the rate performance depends on SNR, precoder design as well as spatial correlation.

Hereafter, we focus on user A and similar results can be derived for user B. The received signal of user A can be written as

\vspace{-10pt}

\begin{equation} \label{eq:samat}
\mathbf{y}_{\scriptscriptstyle{A}} = \mathbf{H}_1 \mathbf{P}^{1/2}_{\scriptscriptstyle{A}} \mathbf{s}_{\scriptscriptstyle{A}} + \mathbf{H}_2 \mathbf{P}^{1/2}_{\scriptscriptstyle{B}} \mathbf{s}_{\scriptscriptstyle{B}} + \mathbf{H}_3 \,\mathbf{s}^p_{\scriptscriptstyle{A}} + \mathbf{H}_4 \,\mathbf{s}^p_{\scriptscriptstyle{B}} + \mathbf{n}_{\scriptscriptstyle{A}},
\end{equation}
where
\begin{small}
\begin{eqnarray} \label{eq:samat_rx2}
\mathbf{H}_1 \triangleq \begin{bmatrix} \mathbf{h}^H_1 \mathbf{W} \\ \mathbf{0} \\ \sqrt{P_{\scriptscriptstyle{8}}} \, h^{\ast}_{\scriptscriptstyle{31}} \mathbf{g}^H_1 \mathbf{W} \end{bmatrix} \mathbf{H}_3 \triangleq \begin{bmatrix} 0 & 0 \\ \sqrt{P_{\scriptscriptstyle{6}}}\, \mathbf{h}^H_2 \mathbf{w}_3 & 0 \\ 0 & \sqrt{P_{\scriptscriptstyle{9}}}\, \mathbf{h}^H_3 \mathbf{w}_3  \end{bmatrix},
\end{eqnarray}
\end{small}

\vspace{-5pt}

\begin{small}
\begin{eqnarray} \label{eq:samat_rx}
\mathbf{H}_2 \triangleq \begin{bmatrix} \mathbf{h}^H_1 \mathbf{Q} \\ \sqrt{P_{\scriptscriptstyle{5}}} \, h^{\ast}_{\scriptscriptstyle{21}} \mathbf{h}^H_1 \mathbf{Q}  \\ \mathbf{0} \end{bmatrix} \mathbf{H}_4 \triangleq \begin{bmatrix} 0 & 0 \\ \sqrt{P_{\scriptscriptstyle{7}}}\, \mathbf{h}^H_2 \mathbf{q}_3 & 0 \\0&\sqrt{P_{\scriptscriptstyle{10}}}\, \mathbf{h}^H_3 \mathbf{q}_3  \end{bmatrix},
\end{eqnarray}
\end{small}and $\mathbf{y}_{\scriptscriptstyle{A}} \triangleq [y_{\scriptscriptstyle{A1}},y_{\scriptscriptstyle{A2}},y_{\scriptscriptstyle{A3}}]^T$, $\mathbf{s}^p_{\scriptscriptstyle{A}} \triangleq [s_{\scriptscriptstyle{A}3}, s_{\scriptscriptstyle{A}4}]^T$,
$\mathbf{s}^p_{\scriptscriptstyle{B}} \triangleq [s_{\scriptscriptstyle{B}3}, s_{\scriptscriptstyle{B}4}]^T$. $\mathbf{n}_{\scriptscriptstyle{A}} \triangleq [n_{\scriptscriptstyle{A1}}, n_{\scriptscriptstyle{A2}}, n_{\scriptscriptstyle{A3}}]^T$ with $n_{\scriptscriptstyle{Aj}} \sim \mathcal{CN}(0, 1)$. The decoding procedure that mainly uses interference alignment and cancellation (similar to \cite{maddah2012}) is described as follows. Denote $\widetilde{\mathbf{y}}_{\scriptscriptstyle{A}}$ as the received signal after subtracting $\sqrt{P_{\scriptscriptstyle{5}}} \, h^{\ast}_{\scriptscriptstyle{21}} \cdot y_{\scriptscriptstyle{A1}}$ from $y_{\scriptscriptstyle{A2}}$ and retaining $y_{\scriptscriptstyle{A1}}$\footnote{A(MAT)-based schemes use one observation to completely remove the overheard interference and two independent observations remain to resolve two symbols. By contrast, we cancel the overheard interference while we maintain all three observations. The reasons are explained as follows: 1) In some cases (e.g., highly correlated channel) where SAMAT boils down to SBF in each time slot, symbols need to be decoded slot by slot. However, conventional decoding strategy causes rate loss because one observation is dropped. 2) The ergodic rate of the proposed decoding method is slightly better than the conventional one, due to one more observation even with strong interference.}. First, decode the private symbols $(\mathbf{s}_{\scriptscriptstyle{A}})$ by regarding the extra symbols ($\mathbf{s}^p_{\scriptscriptstyle{A}}, \mathbf{s}^p_{\scriptscriptstyle{B}}$) as interference:

\vspace{-5pt}

\begin{equation} \label{eq:samat2}
\widetilde{\mathbf{y}}_{\scriptscriptstyle{A}} = \widetilde{\mathbf{H}}_1 \mathbf{P}^{1/2}_{\scriptscriptstyle{A}} \mathbf{s}_{\scriptscriptstyle{A}} + \underbrace{\widetilde{\mathbf{H}}_2 \mathbf{P}^{1/2}_{\scriptscriptstyle{B}} \mathbf{s}_{\scriptscriptstyle{B}} + \mathbf{H}_3 \,\mathbf{s}^p_{\scriptscriptstyle{A}} + \mathbf{H}_4 \,\mathbf{s}^p_{\scriptscriptstyle{B}} + \widetilde{\mathbf{n}}_{\scriptscriptstyle{A}}}_{\mathbf{z}},
\end{equation}
where $\widetilde{\mathbf{H}}_1 = [(\mathbf{h}^H_1 \mathbf{W})^T, -(\sqrt{P_{\scriptscriptstyle{5}}} \, h^{\ast}_{\scriptscriptstyle{21}} \mathbf{h}^H_1 \mathbf{W})^T, (\sqrt{P_{\scriptscriptstyle{8}}} \, h^{\ast}_{\scriptscriptstyle{31}} \mathbf{g}^H_1 \mathbf{W})^T]^T$, $\widetilde{\mathbf{H}}_2 = [(\mathbf{h}^H_1 \mathbf{Q})^T, \; \mathbf{0}^T, \; \mathbf{0}^T]^T$ and $\widetilde{\mathbf{n}}_{\scriptscriptstyle{A}} = [n_{\scriptscriptstyle{A1}}, n_{\scriptscriptstyle{A2}} - \sqrt{P_{\scriptscriptstyle{5}}} \, h^{\ast}_{\scriptscriptstyle{21}} n_{\scriptscriptstyle{A1}}, n_{\scriptscriptstyle{A3}}]^T$. $\mathbf{K}$ is the covariance matrix of the interference plus noise vector $\mathbf{z}$. It is given by $\text{diag} (k_1, k_2, k_3)$, where $k_1 = 1 + |\mathbf{h}^H_1 \mathbf{Q} \mathbf{P}_{\scriptscriptstyle{B}}|^2, k_2 = 1 + P_{\scriptscriptstyle{5}}|h_{\scriptscriptstyle{21}}|^2 + P_{\scriptscriptstyle{6}}|\mathbf{h}^H_2 \mathbf{w}_3|^2 + P_{\scriptscriptstyle{7}}|\mathbf{h}^H_2 \mathbf{q}_3|^2, k_3 = 1 + P_{\scriptscriptstyle{9}}|\mathbf{h}^H_3 \mathbf{w}_3|^2 + P_{\scriptscriptstyle{10}}|\mathbf{h}^H_3 \mathbf{q}_3|^2$. To resolve $\mathbf{s}_{\scriptscriptstyle{A}}$, MMSE-SIC receiver is applied on \eqref{eq:samat2} and the ergodic sum-rate of $\mathbf{s}_{\scriptscriptstyle{A}}$ can then be written as

\begin{equation} \label{eq:smat_amat}
R_{\mathbf{s}_{\scriptscriptstyle{A}}} = E \left[ \log_2  \det\left( \mathbf{I} + \mathbf{P}_{\scriptscriptstyle{A}} \widetilde{\mathbf{H}}_1^H \mathbf{K}^{-1} \widetilde{\mathbf{H}}_1 \mathbf{P}_{\scriptscriptstyle{A}} \right)\right].
\end{equation}

Once $\mathbf{s}_{\scriptscriptstyle{A}}$ is obtained, we can subtract it from $\widetilde{\mathbf{y}}_{\scriptscriptstyle{A}}$. Then, we can decode the extra symbols by taking the second and third entries of $\widetilde{\mathbf{y}}_{\scriptscriptstyle{A}}$ as $\hat{\mathbf{y}}_{\scriptscriptstyle{A}}$:

\begin{equation} \label{eq:rx_sbf}
\hat{\mathbf{y}}_{\scriptscriptstyle{A}} = \hat{\mathbf{H}}_3 \,\mathbf{s}^p_{\scriptscriptstyle{A}} + \hat{\mathbf{H}}_4 \,\mathbf{s}^p_{\scriptscriptstyle{B}} + \hat{\mathbf{n}}_{\scriptscriptstyle{A}},
\end{equation}
where $\hat{\mathbf{H}}_3 = \text{diag}(\sqrt{P_{\scriptscriptstyle{6}}}\, \mathbf{h}^H_2 \mathbf{w}_3, \, \sqrt{P_{\scriptscriptstyle{9}}}\, \mathbf{h}^H_3 \mathbf{w}_3)$, $\hat{\mathbf{H}}_4 = \text{diag}(\sqrt{P_{\scriptscriptstyle{7}}}\, \mathbf{h}^H_2 \mathbf{q}_3, \, \sqrt{P_{\scriptscriptstyle{10}}}\, \mathbf{h}^H_3 \mathbf{q}_3)$ and $\hat{\mathbf{n}}_{\scriptscriptstyle{A}} = [ n_{\scriptscriptstyle{A2}} - \sqrt{P_{\scriptscriptstyle{5}}} \, h^{\ast}_{\scriptscriptstyle{21}} n_{\scriptscriptstyle{A1}}, n_{\scriptscriptstyle{A3}}]^T$. The covariance matrix of $\hat{\mathbf{n}}_{\scriptscriptstyle{A}}$ is given by $\mathbf{N} = \text{diag} (1 + P_{\scriptscriptstyle{5}}|h_{\scriptscriptstyle{21}}|^2, \, 1)$. The ergodic sum-rate of $\mathbf{s}^p_{\scriptscriptstyle{A}}$ is given by

\vspace{-5pt}

\begin{equation} \label{eq:smat_sbf}
R^p_{\mathbf{s}_{\scriptscriptstyle{A}}} = E \left[ \log_2  \det\left( \mathbf{I} + \hat{\mathbf{H}}^H_3 (\mathbf{N} + \hat{\mathbf{H}}_4 \hat{\mathbf{H}}^H_4 )^{-1} \hat{\mathbf{H}}_3 \right)\right].
\end{equation}

It is challenging to obtain the closed-form expression for the ergodic rate. Instead, we derive a tractable approximation and optimize the power allocation based on such approximation.

$\textbf{Proposition 3:}$ The achievable ergodic sum-rate per slot at user A with linear beamforming can be approximated as $R_{\scriptscriptstyle{A}} \triangleq \frac{1}{3} (R_{\mathbf{s}_{\scriptscriptstyle{A}}} + R_{\mathbf{s}^p_{\scriptscriptstyle{A}}} )$ where

\begin{flalign} \label{eq:smat_rate1}
R_{\mathbf{s}_{\scriptscriptstyle{A}}} \approx \log_2 \Big (1 &+ \delta_{\scriptscriptstyle{A1}} \left(\tau_{\scriptscriptstyle{A1}}P_{\scriptscriptstyle{1}} + \tau_{\scriptscriptstyle{A2}}P_{\scriptscriptstyle{2}} \right) + \delta_{\scriptscriptstyle{A2}} \left(\lambda_{\scriptscriptstyle{B1}}P_{\scriptscriptstyle{1}} + \lambda_{\scriptscriptstyle{B2}}P_{\scriptscriptstyle{2}} \right) \nonumber\\
&+ \delta_{\scriptscriptstyle{A1}} \delta_{\scriptscriptstyle{A2}} \Theta_{\scriptscriptstyle{A}} P_{\scriptscriptstyle{1}}P_{\scriptscriptstyle{2}} \Big )
\end{flalign}
\begin{flalign}
R_{\mathbf{s}^p_{\scriptscriptstyle{A}}} \approx \log_2 \left(1 + \frac{\tau_{\scriptscriptstyle{A3}}P_{\scriptscriptstyle{6}}}{1+P_{\scriptscriptstyle{5}}+\lambda_{\scriptscriptstyle{A3}}P_{\scriptscriptstyle{7}}} \right) + \log_2 \left(1 + \frac{\tau_{\scriptscriptstyle{A3}}P_{\scriptscriptstyle{9}}} {1+\lambda_{\scriptscriptstyle{A3}}P_{\scriptscriptstyle{10}}} \right).\nonumber \\
\end{flalign}

Similarly, we have $R_{\scriptscriptstyle{B}}\triangleq\frac{1}{3}(R_{\mathbf{s}_{\scriptscriptstyle{B}}} +R_{\mathbf{s}^p_{\scriptscriptstyle{B}}})$ and
\begin{flalign} \label{eq:smat_rate2}
R_{\mathbf{s}_{\scriptscriptstyle{B}}} \approx \log_2 \Big (1 &+ \delta_{\scriptscriptstyle{B1}} \left(\tau_{\scriptscriptstyle{B1}}P_{\scriptscriptstyle{3}} + \tau_{\scriptscriptstyle{B2}}P_{\scriptscriptstyle{4}} \right) + \delta_{\scriptscriptstyle{B2}} \left(\lambda_{\scriptscriptstyle{A1}}P_{\scriptscriptstyle{3}} + \lambda_{\scriptscriptstyle{A2}}P_{\scriptscriptstyle{4}} \right) \nonumber\\ &+ \delta_{\scriptscriptstyle{B1}}\delta_{\scriptscriptstyle{B2}} \Theta_{\scriptscriptstyle{B}} P_{\scriptscriptstyle{3}}P_{\scriptscriptstyle{4}} \Big )
\end{flalign}

\vspace{-10pt}

\begin{flalign}
R_{\mathbf{s}^p_{\scriptscriptstyle{B}}} \approx \log_2 \left(1 + \frac{\tau_{\scriptscriptstyle{B3}}P_{\scriptscriptstyle{7}}}{1+\lambda_{\scriptscriptstyle{B3}}P_{\scriptscriptstyle{6}}} \right) + \log_2 \left(1 + \frac{\tau_{\scriptscriptstyle{B3}}P_{\scriptscriptstyle{10}}} {1+P_{\scriptscriptstyle{8}}+\lambda_{\scriptscriptstyle{B3}}P_{\scriptscriptstyle{9}}} \right), \nonumber\\ \label{eq:smat_rate22}
\end{flalign}
where \vspace{-10pt}

\begin{flalign} \label{eq:smat_rate3}
\delta_{\scriptscriptstyle{A1}} =& \frac{1}{1+\lambda_{\scriptscriptstyle{A1}}P_{\scriptscriptstyle{3}} + \lambda_{\scriptscriptstyle{A2}}P_{\scriptscriptstyle{4}}} + \frac{P_{\scriptscriptstyle{5}}} {1+P_{\scriptscriptstyle{5}}+ \tau_{\scriptscriptstyle{A3}}P_{\scriptscriptstyle{6}}+ \lambda_{\scriptscriptstyle{A3}}P_{\scriptscriptstyle{7}} } \nonumber \\
\delta_{\scriptscriptstyle{B1}} =& \frac{1}{1+\lambda_{\scriptscriptstyle{B1}}P_{\scriptscriptstyle{1}} + \lambda_{\scriptscriptstyle{B2}}P_{\scriptscriptstyle{2}}} + \frac{P_{\scriptscriptstyle{8}}} {1+P_{\scriptscriptstyle{8}}+ \lambda_{\scriptscriptstyle{B3}}P_{\scriptscriptstyle{9}}+ \tau_{\scriptscriptstyle{B3}}P_{\scriptscriptstyle{10}}} \nonumber \\
\delta_{\scriptscriptstyle{A2}} =& \frac{P_{\scriptscriptstyle{8}}} {1+ \tau_{\scriptscriptstyle{A3}}P_{\scriptscriptstyle{9}}+ \lambda_{\scriptscriptstyle{A3}}P_{\scriptscriptstyle{10}} },
\delta_{\scriptscriptstyle{B2}} = \frac{P_{\scriptscriptstyle{5}}} {1+ \lambda_{\scriptscriptstyle{B3}}P_{\scriptscriptstyle{6}}+ \tau_{\scriptscriptstyle{B3}}P_{\scriptscriptstyle{7}} }
\end{flalign}

\vspace{-7pt}

\begin{flalign} 
\lambda_{\scriptscriptstyle{A1}} &= \mathbf{q}^H_1 \mathbf{R}_{\scriptscriptstyle{A}} \mathbf{q}_1, \; \lambda_{\scriptscriptstyle{A2}} = \mathbf{q}^H_2 \mathbf{R}_{\scriptscriptstyle{A}} \mathbf{q}_2, \; \lambda_{\scriptscriptstyle{B1}} = \mathbf{w}^H_1 \mathbf{R}_{\scriptscriptstyle{B}} \mathbf{w}_1 \label{smat_th3} \nonumber \\
\lambda_{\scriptscriptstyle{B2}} &= \mathbf{w}^H_2 \mathbf{R}_{\scriptscriptstyle{B}} \mathbf{w}_2, \; \tau_{\scriptscriptstyle{A1}} = \mathbf{w}^H_1 \mathbf{R}_{\scriptscriptstyle{A}} \mathbf{w}_1, \; \tau_{\scriptscriptstyle{A2}} = \mathbf{w}^H_2 \mathbf{R}_{\scriptscriptstyle{A}} \mathbf{w}_2 \nonumber \\
\tau_{\scriptscriptstyle{B1}} &= \mathbf{q}^H_1 \mathbf{R}_{\scriptscriptstyle{B}} \mathbf{q}_1, \;\; \tau_{\scriptscriptstyle{B2}} = \mathbf{q}^H_2 \mathbf{R}_{\scriptscriptstyle{B}} \mathbf{q}_2 , \;\; \lambda_{\scriptscriptstyle{A3}} = \mathbf{q}^H_3 \mathbf{R}_{\scriptscriptstyle{A}} \mathbf{q}_3 \nonumber\\
\lambda_{\scriptscriptstyle{B3}} &= \mathbf{w}^H_3 \mathbf{R}_{\scriptscriptstyle{B}} \mathbf{w}_3, \; \tau_{\scriptscriptstyle{A3}} = \mathbf{w}^H_3 \mathbf{R}_{\scriptscriptstyle{A}} \mathbf{w}_3, \; \tau_{\scriptscriptstyle{B3}} = \mathbf{q}^H_3 \mathbf{R}_{\scriptscriptstyle{B}} \mathbf{q}_3
\end{flalign}

\vspace{-12pt}

\begin{small}
\begin{equation}
\Theta_{\scriptscriptstyle{A}} = \text{Tr}(\mathbf{W}^H \mathbf{R}_{\scriptscriptstyle{A}} \mathbf{W}) \text{Tr}(\mathbf{W}^H \mathbf{R}_{\scriptscriptstyle{B}} \mathbf{W}) -  \text{Tr}(\mathbf{W}^H \mathbf{R}_{\scriptscriptstyle{A}} \mathbf{W} \mathbf{W}^H \mathbf{R}_{\scriptscriptstyle{B}} \mathbf{W}) \nonumber
\end{equation}
\end{small}

\vspace{-30pt}

\begin{small}
\begin{equation} \label{eq:smat_rate5}
\hspace{-5pt} \Theta_{\scriptscriptstyle{B}} = \text{Tr}(\mathbf{Q}^H \mathbf{R}_{\scriptscriptstyle{A}} \mathbf{Q}) \text{Tr}(\mathbf{Q}^H \mathbf{R}_{\scriptscriptstyle{B}} \mathbf{Q}) -  \text{Tr}(\mathbf{Q}^H \mathbf{R}_{\scriptscriptstyle{A}} \mathbf{Q} \mathbf{Q}^H \mathbf{R}_{\scriptscriptstyle{B}} \mathbf{Q}).
\end{equation}
\end{small}

\vspace{-10pt}

\begin{IEEEproof}
Refer to Appendix \ref{sec:Proof_props2} for proof.
\end{IEEEproof}

\emph{Remark 2:} Compared to the interference quantization approach in \cite{yang2013}, the analog transmission induces a noise enhancement. Namely, interference alignment cancels the overheard interference while scaling up the noise by $P_{\scriptscriptstyle{5}}$ $(P_{\scriptscriptstyle{8}})$. This noise enhancement can be observed in \eqref{eq:samat2} $\thicksim$ \eqref{eq:smat_rate22}. At low SNR, the proposed SAMAT scheme behaves as SBF in each time slot. The scaling factors are small and therefore the effect of noise enhancement is negligible. The gain over AMAT mainly comes from extra symbol transmission and statistical precoding. At high SNR, the proposed SAMAT scheme behaves as AMAT, achieving a DoF of $\frac{4}{3}$. In this case, the ergodic rates of extra symbols can be eliminated by noise enhancement. Namely, we have little benefit by transmitting extra symbols. However, the proposed SAMAT scheme still achieves significant gain over SBF and AMAT by power allocation optimization and statistical precoding.

With predefined beamforming vectors, the proposed SAMAT scheme softly bridges between SBF and AMAT by power control. Let us concentrate on two cases:

\textbf{\emph{case 1}}: bridge between SWEBF and AMAT, $\mathbf{w}_1 = \mathbf{u}_{\text{max}}(\mathbf{R}_{\scriptscriptstyle{B}}), \, \mathbf{q}_1 = \mathbf{u}_{\text{max}} (\mathbf{R}_{\scriptscriptstyle{A}}), \, \mathbf{w}_2 = \mathbf{w}_3 = \mathbf{w}_{\scriptscriptstyle{WE}} = \mathbf{u}_{\text{min}}(\mathbf{R}_{\scriptscriptstyle{B}}), \, \mathbf{q}_2 = \mathbf{q}_3 = \mathbf{q}_{\scriptscriptstyle{WE}} = \mathbf{u}_{\text{min}}(\mathbf{R}_{\scriptscriptstyle{A}})$;

\textbf{\emph{case 2}}: bridge between SGEBF and AMAT, $\mathbf{w}_1 = \mathbf{u}_\text{min}(\mathbf{R}^{-1}_{\scriptscriptstyle{B}} \mathbf{R}_{\scriptscriptstyle{A}}), \, \mathbf{q}_1 = \mathbf{u}_\text{min}(\mathbf{R}^{-1}_{\scriptscriptstyle{A}} \mathbf{R}_{\scriptscriptstyle{B}}), \, \mathbf{w}_2 = \mathbf{w}_3 = \mathbf{w}_{\scriptscriptstyle{GE}} = \mathbf{u}_\text{max}(\mathbf{R}^{-1}_{\scriptscriptstyle{B}} \mathbf{R}_{\scriptscriptstyle{A}}), \, \mathbf{q}_2 = \mathbf{q}_3 = \mathbf{q}_{\scriptscriptstyle{GE}} = \mathbf{u}_\text{max}(\mathbf{R}^{-1}_{\scriptscriptstyle{A}} \mathbf{R}_{\scriptscriptstyle{B}})$.

\textbf{\emph{case 1}} is used to show the efficacy of the power allocation optimization technique by which the proposed SAMAT scheme can softly bridge between SWEBF and AMAT. Beyond this, \textbf{\emph{case 2}} makes better use of statistical CSIT in the sense that SGEBF exhibits higher robustness compared to SWEBF. Instead of using the optimized AMAT precoders in transmission stage I ($\mathbf{w}_1, \mathbf{w}_2, \mathbf{q}_1, \mathbf{q}_2$ as developed in Section \ref{sec:AMAT}), we use the precoders above ($\mathbf{w}_1, \mathbf{w}_2, \mathbf{q}_1, \mathbf{q}_2$ as WE/GE precoder) and the motivations are explained as follows. First, the optimal precoders in Section \ref{sec:AMAT} that maximize the ergodic sum-rate of AMAT under equal power allocation are not necessarily optimal for SAMAT with power control. Second, SAMAT boils down to SBF at low to intermediate SNR in highly correlated channel, where the optimized AMAT precoders may cause a poorer rate performance compared to the WE/GE precoders. In order to softly bridge between SBF and AMAT, we adopt the precoder design as above.

Consequently, $R_{\scriptstyle{\text{sum}}} \triangleq R_{\scriptscriptstyle{A}} + R_{\scriptscriptstyle{B}}$ and the ergodic sum-rate optimization problem is formulated as:

\vspace{-9pt}

\begin{equation} \label{eq:opt_prob}
\begin{array}{lcl}
\underset{\{P_i\}}{\text{max}} \; R_{\scriptstyle{\text{sum}}} \quad \text{s.t.} \quad \bar{P}_{\text{c}} = 3P,\; P_i \ge 0 \quad i = 1,\ldots,10.
\end{array}
\end{equation}

It was shown that the sum-rate optimization generally leads to an intractable NP hard problem \cite{chiang2007}. Thus, an algorithm achieving global optimum cannot be expected. However, Sequential Quadratic Programming (SQP) algorithm provides an efficient way to solve non-linear constrained optimization problem. An overview on SQP is provided in \cite{boggs1995,fletcher2010,morales2012}. Briefly, a quadratic approximation of the Lagrangian function is made by applying quasi-Newton updating method. The consequent QP subproblem can be optimally solved and then the solution is used as a search direction. With proper line search, an estimate of the solution is computed for the next iteration. This SQP algorithm can guarantee a super-linear convergence to a local minimum.

In order to get insights into the optimal power allocation, a necessary condition for optimality of the constrained problem \eqref{eq:opt_prob} is identified from the Karush-Kuhn-Tucker (KKT) equations. Thus, the optimum power allocation that depends on the precoders and the channel covariance matrices is stated as follows.

$\textbf{Theorem 2:}$ At high SNR, the optimal power allocation that maximizes $R_{\scriptstyle{\text{sum}}}$ in \eqref{eq:opt_prob} satisfies:

\begin{equation} \label{eq:th3}
\frac{ P_{\scriptscriptstyle{1}} }{P_{\scriptscriptstyle{2}}} = \frac{1+\lambda_{\scriptscriptstyle{B2}} P_{\scriptscriptstyle{8}} }{1+\lambda_{\scriptscriptstyle{B1}} P_{\scriptscriptstyle{8}}}, \qquad \frac{ P_{\scriptscriptstyle{3}} }{P_{\scriptscriptstyle{4}}} = \frac{1+\lambda_{\scriptscriptstyle{A2}} P_{\scriptscriptstyle{5}} }{1+\lambda_{\scriptscriptstyle{A1}} P_{\scriptscriptstyle{5}}}
\end{equation}
where $\lambda_{\scriptscriptstyle{A1}}, \lambda_{\scriptscriptstyle{A2}}, \lambda_{\scriptscriptstyle{B1}}, \lambda_{\scriptscriptstyle{B2}}$ are defined in \eqref{smat_th3}.

\begin{IEEEproof}
At high SNR, problem \eqref{eq:opt_prob} can be rewritten as
\begin{equation} \label{eq:opt_prob2}
\left\{
\begin{array}{lcl}
\underset{\{P_i\}}{\text{max}} \; R_{\scriptstyle{\text{sum}}} \overset{{(a)}}{\approx} \log_2 \left( \delta_{\scriptscriptstyle{A1}}\delta_{\scriptscriptstyle{A2}} \Theta_{\scriptscriptstyle{A}} P_{\scriptscriptstyle{1}}P_{\scriptscriptstyle{2}} \right) + \log_2 \left(\delta_{\scriptscriptstyle{B1}}\delta_{\scriptscriptstyle{B2}} \Theta_{\scriptscriptstyle{B}} P_{\scriptscriptstyle{3}} P_{\scriptscriptstyle{4}} \right) \\
\qquad \quad  + \; \log_2 \left(1 + \frac{\tau_{\scriptscriptstyle{A3}}P_{\scriptscriptstyle{6}}}{1+P_{\scriptscriptstyle{5}}+\lambda_{\scriptscriptstyle{A3}}P_{\scriptscriptstyle{7}}} \right) + \log_2 \left(1 + \frac{\tau_{\scriptscriptstyle{A3}}P_{\scriptscriptstyle{9}}}{1+\lambda_{\scriptscriptstyle{A3}}P_{\scriptscriptstyle{10}}}\right)  \\
\qquad \quad  + \; \log_2 \left(1 + \frac{\tau_{\scriptscriptstyle{B3}}P_{\scriptscriptstyle{7}}}{1+\lambda_{\scriptscriptstyle{B3}}P_{\scriptscriptstyle{6}}} \right) + \log_2 \left(1 + \frac{\tau_{\scriptscriptstyle{B3}}P_{\scriptscriptstyle{10}}} {1+P_{\scriptscriptstyle{8}}+\lambda_{\scriptscriptstyle{B3}}P_{\scriptscriptstyle{9}}} \right)\\
\text{s.t.} \quad \bar{P}_{\text{c}} - 3P = 0, \; P_i \ge 0 \quad i = 1,\ldots,10
\end{array}
\right.
\end{equation}
where $\overset{{(a)}}{\approx}$ comes from the fact that the last terms in \eqref{eq:smat_rate1} and \eqref{eq:smat_rate2} are dominant at high SNR. Based on KKT necessary conditions, there exist multipliers $\lambda \text{ and } \mu_{1}, \mu_{2}$ such that

\begin{equation} \label{eq:opt_kkt}
\left\{
\begin{array}{lcl}
\nabla R_{\scriptstyle{\text{sum}}}(P_{\scriptscriptstyle{1}}) = \lambda (1+ \lambda_{\scriptscriptstyle{B1}} P_{\scriptscriptstyle{8}}) + \mu_{1} \\
\nabla R_{\scriptstyle{\text{sum}}}(P_{\scriptscriptstyle{2}}) = \lambda (1+ \lambda_{\scriptscriptstyle{B2}} P_{\scriptscriptstyle{8}}) + \mu_{2} \\
\mu_{1} P_{\scriptscriptstyle{1}} = 0, \; \mu_{2} P_{\scriptscriptstyle{2}} = 0,
\end{array}
\right.
\end{equation}
where $P_{\scriptscriptstyle{1}} \neq 0, P_{\scriptscriptstyle{2}} \neq 0$, otherwise DoF loss occurs due to $R_{\mathbf{s}_{\scriptscriptstyle{A}}} = 0$. Therefore, we have $\mu_{1} = \mu_{2} = 0$ and the first equation in \eqref{eq:th3} can be computed from \eqref{eq:opt_kkt}. Likewise, the second equation can be obtained.\end{IEEEproof}

\emph{Remark 3:} As can be seen from \eqref{eq:th3}, the power allocation depends on the spatial correlation, the precoder design as well as SNR. Take \textbf{\emph{case 1}} as an example, $\lambda_{\scriptscriptstyle{B2}} = \mathbf{w}^H_2 \mathbf{R}_{\scriptscriptstyle{B}} \mathbf{w}_2 = \lambda_{\text{min}}(\mathbf{R}_{\scriptscriptstyle{B}})$ while $\lambda_{\scriptscriptstyle{B1}} = \lambda_{\text{max}} (\mathbf{R}_{\scriptscriptstyle{B}})$. Then, $\frac{1+\lambda_{\scriptscriptstyle{B2}} P_{\scriptscriptstyle{8}} }{1+\lambda_{\scriptscriptstyle{B1}} P_{\scriptscriptstyle{8}}} \le 1$ implies that $P_{\scriptscriptstyle{1}} \le P_{\scriptscriptstyle{2}}$ and likewise $P_{\scriptscriptstyle{3}} \le P_{\scriptscriptstyle{4}}$. This implies that more power needs to be allocated on the weaker eigen-mode ($\mathbf{w}_2$) to constrain the interference imposed to the other desired symbol of the same user. As mentioned before, the transmitted power in $2^{\text{nd}}$ and $3^{\text{rd}}$ time slots is a function of beamformers. Such power allocation method enables to compress the interference and makes room for delivering two more private symbols. Moreover, consider i.i.d Rayleigh fading channels where $\mathbf{R}_{\scriptscriptstyle{A}} = \mathbf{R}_{\scriptscriptstyle{B}} = \mathbf{I}$. $\lambda_{\scriptscriptstyle{B2}} = \mathbf{w}^H_2 \mathbf{R}_{\scriptscriptstyle{B}} \mathbf{w}_2 = 1$ and likewise $\lambda_{\scriptscriptstyle{B1}} = \lambda_{\scriptscriptstyle{A1}} = \lambda_{\scriptscriptstyle{A2}} = 1$. Based on \eqref{eq:th3} and symmetry, the optimal power allocation satisfies $P_{\scriptscriptstyle{1}} = P_{\scriptscriptstyle{2}} = P_{\scriptscriptstyle{3}} = P_{\scriptscriptstyle{4}}$ and therefore SAMAT boils down to AMAT. It makes sense because no correlated information is available to suppress the interference. In this case, equal power allocation is the optimal choice.

To operate the proposed SAMAT transmission protocol, the signaling and feedback procedure is described as follows. Using LTE-A framework \cite{Bocc2012}, channel state information reference signals (CSI-RS) are transmitted to enable the receiver to measure the short-term CSI and the long-term CSI (channel covariance matrix), which are then fed back to the transmitter via a delayed but assumed perfect feedback link. The long-term CSI only varies at a very slow pace and is therefore not affected by the delay. However by the time the transmitter has acquired the short term CSI, the channel has changed and the transmitter only has knowledge of a completely stale short-term CSIT. Based on the long-term and the short-term CSIT, the transmitter computes the precoders and the power allocation and constructs the transmitted signals that are then transmitted using demodulation reference signals (DM-RS) \cite{Lim2013}. As far as the implementation complexity is concerned, the potential challenge of SAMAT lies in numerical power allocation computation. Eq. \eqref{eq:th3} provides a necessary condition which reveals the basic relationship between the power allocation and the spatial correlation matrices as well as the precoders. It would be helpful to identify a sufficient condition and further closed-form power allocation.

\section{PERFORMANCE EVALUATION} \label{numresults}
We provide numerical results to show the efficacy of the proposed precoder design and power allocation strategy. A single parameter exponential correlation model \cite{clerckx2008} is considered as

\begin{equation} \label{R}
\mathbf{R}_k = \begin{bmatrix} 1 & t_k & \ldots & t^{M-1}_k  \\ t^H_k & 1 & \ldots & t^{M-2}_k \\ \vdots & & \ddots & \\ (t^H_k)^{M-1} & \ldots & t^H_k & 1  \end{bmatrix}
\end{equation}
where $t_k$ denotes the transmit correlation coefficient $t_k = |t_k|\, e^{j\phi_{\scriptscriptstyle{k}}}, \, \phi_{\scriptscriptstyle{k}} \in [0, \, 2\pi], \; k = A, B$. Throughout the paper, we use high(low) correlation to indicate large(small) condition number of the spatial correlation matrix, which corresponds to a large(small) $|t_k|$ in the exponential model. A large family of spatial correlation is tested to verify our analysis. With the help of the optimization tool in Matlab, `fmincon' is used to implement the SQP algorithm.

\begin{figure}[t]
  \centering
  \includegraphics[width=3.4in, height = 2.4in]{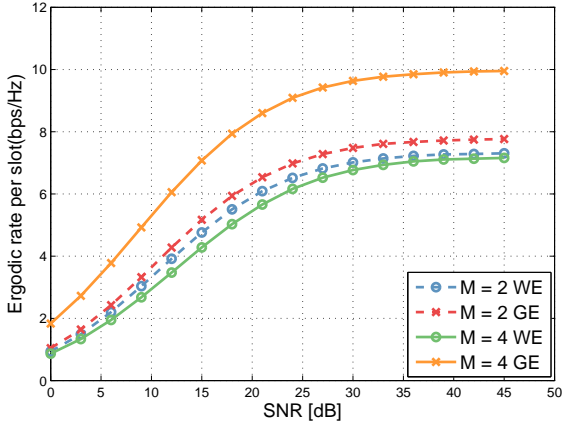}\\
  \caption{Rate performance comparison of SBF with WE and GE precoder.}\label{fig_GEZF}
\end{figure}

\subsection{Precoders Comparison for Statistical Beamforming}
In Fig. \ref{fig_GEZF}, we plot the ergodic sum-rate of SBF with WE/GE precoders, averaged over the randomness in the channels and $\phi_{\scriptscriptstyle{k}}$. The amplitudes of channel correlation coefficients of both users are given by $|t_{\scriptscriptstyle{A}}| = 0.95, \; |t_{\scriptscriptstyle{B}}| = 0.9$ and the superiority of GE over WE precoder is illustrated by two cases ($M = 2,4$). In Fig. \ref{fig_GEZF}, GE beamforming vector shows robustness for large $M$ as well as varying scattering environment (i.e., $\phi_{\scriptscriptstyle{k}}$). Interestingly, WE precoded SBF performs even worse for larger $M$, which is inherently caused by the idea of zero forcing. The precoder is designed to reduce the interference imposed to the unintended user, but may cancel out the desired signal of the intended user. In other words, as $M$ increases, the $M \times 1$ WE precoder $\mathbf{w}$ = $\mathbf{u}_{\text{min}}(\mathbf{R}_{\scriptscriptstyle{B}})$ may fall into the ($M - 1$) dimensional $\mathbb{N}(\mathbf{h})$ with higher probability.

\begin{figure}[t] \centering
\subfigure[$M = 4$.] { \label{fig_AMATth2}
\includegraphics[width=3.4in, height = 2.4in]{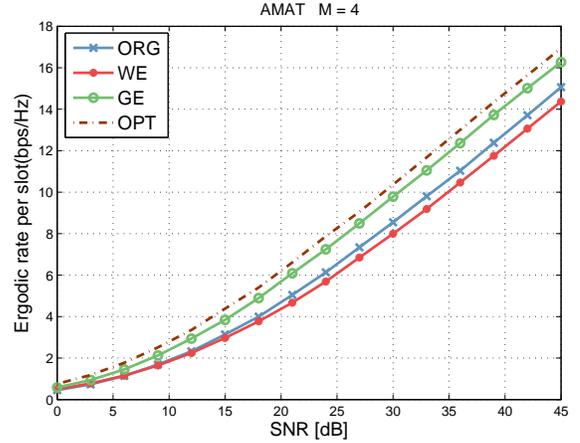}}
\subfigure[$M = 2$.] { \label{fig_AMATco2}
\includegraphics[width=3.4in, height = 2.4in]{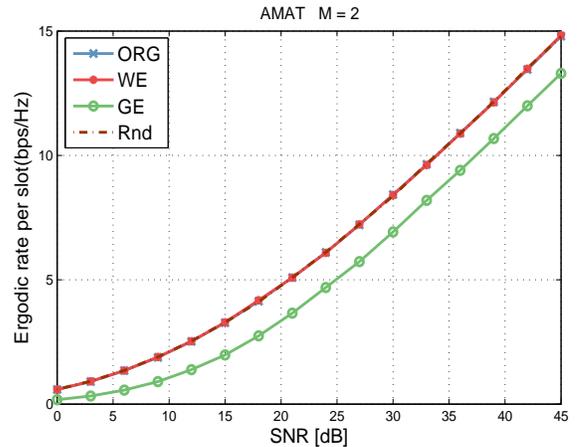}}
\caption{Rate performance comparison of AMAT with different precoders.}
\label{fig_AMATcon}
\end{figure}

\subsection{Precoders Comparison for Alternative MAT}
In Fig. \ref{fig_AMATth2}, we compare the ergodic sum-rate performance of AMAT with different precoding methods. $t_k$ is randomly generated: $|t_k| \in U(0,1), \phi_{\scriptscriptstyle{k}} \in U(0, 2\pi)$. ORG denotes original AMAT that the transmitter sends symbols simply using 2 out of $M$ antennas. WE and GE are statistical precoders defined in section \ref{sec:SBF}. The optimal precoders (OPT) is computed by the proposed iterative algorithm in Table I. We can observe that OPT achieves a better ergodic sum-rate than others.

Meanwhile, Fig. \ref{fig_AMATco2} confirms the validity of proposition 2. It can be seen that any orthogonal beamforming vectors constituting a unitary matrix are optimal for $M = 2$ case. More specifically, Rnd indicates that $\mathbf{W}$ and $\mathbf{Q}$ are randomly generated unitary matrices. WE precoders, corresponding to $\mathbf{u}_{\text{max}}(\mathbf{R}_{k})$ and $\mathbf{u}_{\text{min}}(\mathbf{R}_{k}) \; k = A, B$, also form unitary matrices. ORG becomes an $2\times2$ identity matrix. All these precoders show optimality in terms of the ergodic sum-rate whereas GE does not, because GE precoders fail to form a unitary matrix (since either $\mathbf{R}^{-1}_{\scriptscriptstyle{A}} \mathbf{R}_{\scriptscriptstyle{B}}$ or $\mathbf{R}^{-1}_{\scriptscriptstyle{B}} \mathbf{R}_{\scriptscriptstyle{A}}$ is a normal matrix).

\subsection{Performance of the Proposed Statistical AMAT (SAMAT)}
\begin{figure*}[t] \centering
\subfigure[$|\phi_{\scriptscriptstyle{A}} - \phi_{\scriptscriptstyle{B}}| \ge \frac{\pi}{2}$, SNR $= 20$ dB, $M = 2$.] { \label{fig:SAMAT1}
\includegraphics[width=3.4in, height = 2.4in]{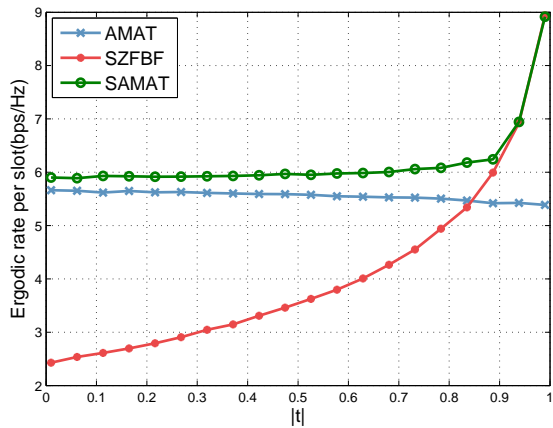}}
\subfigure[$|\phi_{\scriptscriptstyle{A}} - \phi_{\scriptscriptstyle{B}}| \ge \frac{\pi}{2}$, $|t_{\scriptscriptstyle{A}}| = 0.95$,
$|t_{\scriptscriptstyle{B}}| = 0.9$, $M = 2$.] { \label{fig:SAMAT2}
\includegraphics[width=3.4in, height = 2.4in]{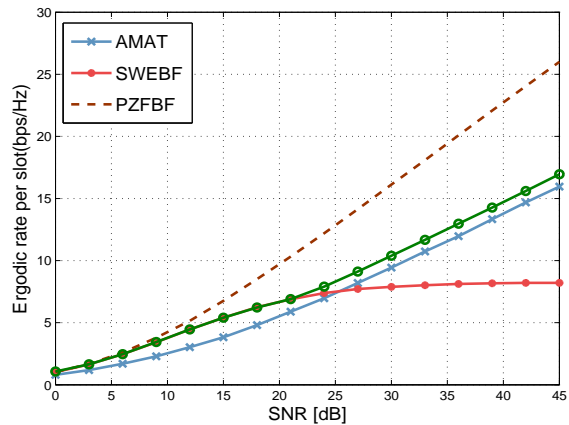}}
\caption{Comparison of the ergodic sum-rate vs. $|t_k|$ or SNR for various schemes.}
\label{fig_SAMAT}
\end{figure*}

Fig. \ref{fig:SAMAT1} depicts the achievable ergodic sum-rate performances of various schemes with two transmit antennas $(M = 2)$: original AMAT, SBF with WE precoders and the proposed SAMAT (\textbf{\emph{case 1}}). We set $|t_{\scriptscriptstyle{A}}| = |t_{\scriptscriptstyle{B}}| = |t|$ that varies between 0 and 1, i.e., from uncorrelated to highly correlated channels. Furthermore, $\phi_{\scriptscriptstyle{A}}$, $\phi_{\scriptscriptstyle{B}}$ are randomly generated with $|\phi_{\scriptscriptstyle{A}} - \phi_{\scriptscriptstyle{B}}| \ge \frac{\pi}{2}$ and SNR $= 20$ dB. As $|t|$ increases, the sum-rate of SBF gradually goes up while a sharp rise occurs at very high correlation level. Because in highly correlated spatial channels, linear beamforming based on statistical information keeps the remaining interference small enough. A special case is the fully correlated channel. When $|t| = 1$, the overheard interference can be completely cancelled.

Moreover, the rate performance of original AMAT also depends on the transmit correlation of the channel. In $M = 2$ case, ORG precoders for original AMAT become $\mathbf{W} = \mathbf{Q} = \mathbf{I}_{2 \times 2}$. Observe in \eqref{eq:amat_rate_appr} and \eqref{eq:theta} that the ergodic rate is a function of $\mathbf{R}_{\scriptscriptstyle{A}}$ and $\mathbf{R}_{\scriptscriptstyle{B}}$. More specifically, $\Theta_{\scriptscriptstyle{A}} = \Theta_{\scriptscriptstyle{B}} = \text{Tr}(\mathbf{R}_{\scriptscriptstyle{A}}) \text{Tr}(\mathbf{R}_{\scriptscriptstyle{B}}) - \text{Tr} (\mathbf{R}_{\scriptscriptstyle{A}} \mathbf{R}_{\scriptscriptstyle{B}})$. With the correlation model in \eqref{R} and the specific phases $\phi_{\scriptscriptstyle{A}}$, $\phi_{\scriptscriptstyle{B}}$, a positive/negative impact of transmit correlation amplitude $|t|$ can be easily computed: $\Theta_{\scriptscriptstyle{A}} = \Theta_{\scriptscriptstyle{B}} = 2\big(1 - |t|^2| \cdot \cos(|\phi_{\scriptscriptstyle{A}} - \phi_{\scriptscriptstyle{B}}|)\big)$. As $|t|$ increases, the transmit correlation is beneficial when $|\phi_{\scriptscriptstyle{A}} - \phi_{\scriptscriptstyle{B}}| > \frac{\pi}{2}$ while it is detrimental when $|\phi_{\scriptscriptstyle{A}} - \phi_{\scriptscriptstyle{B}}| < \frac{\pi}{2}$. When $|\phi_{\scriptscriptstyle{A}} - \phi_{\scriptscriptstyle{B}}| = \frac{\pi}{2}$, the ergodic rate keeps constant irrespectively of $|t|$.

The cross point between SWEBF and AMAT is determined by the spatial correlation level and SNR. Fig. \ref{fig:SAMAT1} reveals that the proposed SAMAT scheme obtains strictly higher rate than SWEBF and AMAT by exploiting both statistical- and delayed-CSIT. Since the channel statistics includes partial current CSI, we can `virtually' regard it as an imperfect current CSI. Fig. \ref{fig:SAMAT1} coincides with Fig. 1 in \cite{yang2013} in the sense that the proposed schemes softly bridge between SWEBF (PZFBF) and AMAT (MAT) in terms of the ergodic sum-rate (DoF).

In addition, for given channel covariance matrices, the ergodic sum-rate of these schemes can be plotted versus SNR. It can be observed from Fig. \ref{fig:SAMAT2} that SAMAT achieves higher rate than SWEBF as well as AMAT along the entire SNR region. It acts as SWEBF at low SNR while it utilizes the DoF capability of AMAT in the high SNR regime. As a comparison, PZFBF with perfect instantaneous CSIT reaches a sum DoF of 2 at high SNR. A variety of spatial correlation is simulated but omitted here for conciseness. Briefly, when $|t_{\scriptscriptstyle{A}}|, |t_{\scriptscriptstyle{B}}| \longrightarrow 0$, the SAMAT transmission protocol boils down to AMAT since no correlated channel information can be exploited to enhance the rate performance. Consider the other extreme $|t_{\scriptscriptstyle{A}}|, |t_{\scriptscriptstyle{B}}| \longrightarrow 1$ but $|\phi_{\scriptscriptstyle{A}} - \phi_{\scriptscriptstyle{B}}| \longrightarrow 0$, it indicates highly correlated channels but their weakest eigen-modes lie in the similar direction. The rate performance of SWEBF is unfavorable and therefore SAMAT also behaves as AMAT.

\begin{figure}[t]
\centering
\includegraphics[width=3.4in, height = 2.4in]{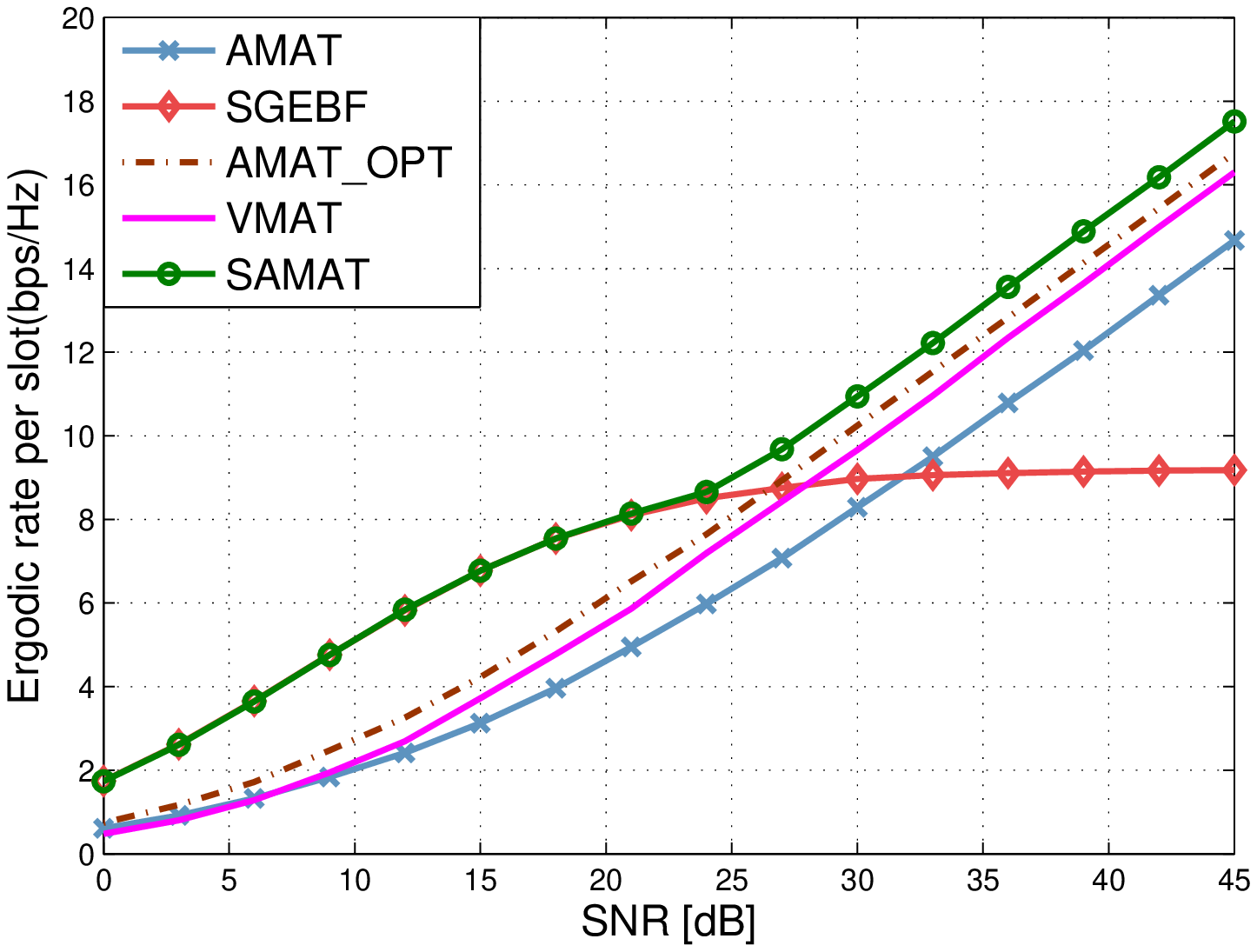}\\
\caption{Comparison of the ergodic sum-rate vs. SNR between SAMAT and baselines, $|t_{\scriptscriptstyle{A}}| = 0.95$, $|t_{\scriptscriptstyle{B}}| = 0.9$, $\phi_{\scriptscriptstyle{A}}, \phi_{\scriptscriptstyle{B}} \in U(0, 2\pi)$, $M = 4$.} \label{fig:SAMAT3}
\end{figure}

Fig. \ref{fig:SAMAT3} illustrates the benefits of the proposed scheme with the power allocation optimization. The transmitter antennas $M = 4$ and robust GE precoders are considered. $\phi_{\scriptscriptstyle{A}}$ and $\phi_{\scriptscriptstyle{B}}$ are randomly generated. Specifically, AMAT indicates the original AMAT with equal power allocation only exploiting delayed CSIT while AMAT\_OPT denotes AMAT precoded by optimal linear beamforming vectors developed in section \ref{sec:AMAT}. SGEBF denotes the SBF scheme with GE precoders. We plot SAMAT (\textbf{\emph{case 2}}) with the SQP algorithm. Moreover, we compare the proposed SAMAT scheme with VMAT \cite{wang2013}. As mentioned before, the power constraint of VMAT in stage II was released. To make a fair comparison, we also apply the long-term power constraint for VMAT and scale it down to $3P$.

In Fig. \ref{fig:SAMAT3}, we observe that AMAT\_OPT enables around 5 dB enhancement over original AMAT at high SNR. VMAT achieves almost the same ergodic sum-rate as AMAT\_OPT, since both schemes exploit statistical CSIT under equal power allocation. However, the SBF scheme still outperforms all of them in a certain range of low to intermediate SNR. The proposed SAMAT framework which is precoded by GE with closed-form power allocation outperforms all these schemes. Meanwhile, with the optimized power allocation computed by the SQP algorithm, the SAMAT scheme maximizes the ergodic sum-rate (further 2 dB over AMAT\_OPT). The enhancement over VAMT/AMAT\_OPT mainly comes from the optimized power allocation.

To sum up, SAMAT boils down to AMAT in low-correlated/uncorrelated channels while for highly correlated scenario where SBF outperforms AMAT, it behaves as SBF in the low to mediate SNR regime and as AMAT at high SNR. In addition, the optimized power values satisfy Theorem 2.

\section{CONCLUSION} \label{conclusion}

This paper aimed to exploit both statistical and outdated CSIT in a MISO broadcast setting to maximize the ergodic sum-rate. We considered the robust design of statistical beamforming vectors for arbitrary transmit antennas, showing the optimality of dominant generalized eigenvectors in maximizing a lower bound of the ergodic sum-rate. Moreover, the optimal precoders were designed to maximize the rate approximation of AMAT under equal power allocation. An iterative algorithm was explored to compute these precoders with fast convergence.

The SBF and AMAT schemes show different performance behaviour depending on the spatial correlation and SNR. To overcome this, the SAMAT transmission protocol was proposed to bridge between SBF and AMAT for a wide range of SNR and an arbitrary number of transmit antennas. In low correlated channel, the SAMAT scheme boils down to AMAT because limited spatial correlation can be exploited to enhance the ergodic sum-rate. For highly correlated scenario, it employs the advantage of SBF in the low to intermediate SNR region and the DoF capability of AMAT at high SNR. To sum up, the proposed SAMAT scheme yields a significant ergodic sum-rate enhancement over both SBF and AMAT. At low SNR, the gain mostly comes from extra symbols transmission. At high SNR, it is achieved by power allocation optimization and statistical precoding.

Numerical results were provided to confirm the design and the analysis of this paper. Simulation outputs illustrated that the proposed SAMAT scheme with optimized power allocation achieves a significant ergodic sum-rate enhancement over both SBF and original AMAT. At last, a joint optimization on precoder design and power allocation is an ongoing work.

\appendices

\section{USEFUL LEMMAS} \label{sec:lemmas}

$\textbf{Lemma 1:}$ Consider a non-zero vector $\mathbf{w} \in \mathbb{C}^M $ and $\mathbf{h}=\mathbf{R}^{1/2} \, \mathbf{h}_{w}$, where $\mathbf{R} \in \Phi_{\text{PD}}$ is a $M\times M$ Hermitian matrix. Then,

\begin{equation} \label{eq:lemma1_1}
E \left[ \ln \left(| \mathbf{h}^H \mathbf{w} |^2\right)\right] = \ln \left(\mathbf{w}^H \mathbf{R} \mathbf{w}\right) - \gamma,
\end{equation}
where $\gamma$ is the Euler constant.

\begin{IEEEproof}
Define $\mathbf{X} \triangleq \mathbf{R}^{1/2} \mathbf{w} \mathbf{w}^H \mathbf{R}^{1/2}$ and decompose it as $\mathbf{X} = \mathbf{U}_{\scriptscriptstyle{X}} \mathbf{\Lambda}_{\scriptscriptstyle{X}} \mathbf{U}^H_{\scriptscriptstyle{X}}$. Due to rank$(\mathbf{X}) = 1$, the diagonal matrix $\mathbf{\Lambda}_{\scriptscriptstyle{X}}$ has only one non-zero entry. Let us define it as the $m$-th entry, denoted by $\lambda_{\scriptscriptstyle{X}}$.

\begin{equation} \label{eq:lemma1_2}
\lambda_{\scriptscriptstyle{X}} =  \text{Tr}(\mathbf{\Lambda}_{\scriptscriptstyle{X}}) \overset{(a)}{=} \text{Tr}(\mathbf{X})  \overset{(b)}{=} \mathbf{w}^H \mathbf{R} \mathbf{w}.
\end{equation}

Equalities $(a)$ and $(b)$ can be easily obtained by applying $\text{Tr}(\mathbf{A}\mathbf{B}) = \text{Tr}(\mathbf{B}\mathbf{A})$. Then

\vspace{-5pt}

\begin{eqnarray}
\label{eq:th1_proof2}
E \left[ \ln \left(| \mathbf{h}^H \mathbf{w} |^2\right)\right]  &=& E \left[ \ln \left(\mathbf{h}^H_w \mathbf{R}^{1/2} \mathbf{w} \mathbf{w}^H \mathbf{R}^{1/2} \mathbf{h}_w  \right) \right] \nonumber\\
&\overset{d}{=}& E \left[ \ln \left(\mathbf{h}^H_w \mathbf{\Lambda}_{\scriptscriptstyle{X}} \mathbf{h}_w \right) \right] \label{lemma1_4} \\
&=& E \left[ \ln \left(\lambda_{\scriptscriptstyle{X}} |h_{w,m}|^2  \right) \right] \label{lemma1_5} 
\end{eqnarray}
where $\overset{d}{=}$ indicates the equivalence in distribution and (\ref{lemma1_5}) is calculated with the non-zero element in $\mathbf{\Lambda}_{\scriptscriptstyle{X}}$. Then, \eqref{eq:lemma1_1} can be obtained via \eqref{eq:lemma1_2} and the fact that $|h_{w,m}|^2 \sim \text{Exp}(1)$.
\end{IEEEproof}

$\textbf{Lemma 2:}$ Suppose $x, y$ are two random variables. $E(y) \neq 0$, let $f(x,y)=\frac{x}{y}$ and $\mathbf{\mu}=(E(x),E(y))=(\mu_x,\mu_y)$. The first order approximation of the expectation of $f(x,y)$ can be written as:

\begin{equation}\label{taylor_exp1}
E \left(\frac{x}{y} \right) = \frac{\mu_x}{\mu_y} + O \bigg(\frac{\text{var(y)} \mu_x }{\mu_y^3} - \frac{\text{cov(x,y)}}{\mu_y^2} \bigg).
\end{equation}

\hspace{10pt}\emph{Proof \,Sketch:} The closed-form of $E \left(\frac{x}{y} \right)$ is unknown, however, it can be calculated via bivariate Taylor expansion at $\mathbf{\mu}$:

\vspace{-10pt}

\begin{eqnarray} \label{taylor_exp2}
E(f(x,y)) &=& \frac{\mu_x}{\mu_y} + \sum^\infty_{n=1} (-1)^n \frac{\mu_x \cdot \pi_{\scriptscriptstyle{0,n}} + \pi_{\scriptscriptstyle{1,n}}}{\mu_y^{n+1}},
\end{eqnarray}
where $\pi_{\scriptscriptstyle{i,j}} = E \left[ (x-\mu_x)^i \cdot (y-\mu_y)^j\right]$. Take the first order approximation of \eqref{taylor_exp2} and \eqref{taylor_exp1} is obtained. Similar results were derived in an alternative manner \cite{rice2009}. However, it is difficult to calculate the high-order terms in \eqref{taylor_exp2} so that the first and second order approximations were used in \cite{HSeltman2012}. It is assumed here that $E \left(\frac{x}{y} \right)$ is bounded and its Taylor expansion converges. Moreover, if $x,y$ are mutually independent nonnegative random variables, the first order approximation is a lower bound, i.e., $E (\frac{x}{y}) \ge \frac{\mu_x}{\mu_y}$.

\section{PROOF OF THEOREM 1} \label{sec:proof_th1}
The proof relies on deriving a lower bound on the achievable ergodic sum-rate. According to \eqref{eq:ergodic_rate}, we can rewrite the ergodic sum-rate as

\begin{equation}
\begin{array}{l}
\hspace{-6pt} R_{\scriptstyle{\text{sum}}}=E \left[\log_2\left(1 + \frac{\rho | \mathbf{h}^H \mathbf{w} |^2}{1 +  \rho | \mathbf{h}^H \mathbf{q} |^2}\right)\right] + E \left[\log_2\left(1 + \frac{\rho | \mathbf{g}^H \mathbf{q} |^2}{1 +  \rho | \mathbf{g}^H \mathbf{w} |^2}\right)\right] \nonumber
\end{array}
\end{equation}

\vspace{-3pt}

\begin{equation}
\begin{array}{l}
\hspace{-30pt} = E \left[\log_2\left(1 + \exp\left(\ln\left(\frac{\rho | \mathbf{h}^H \mathbf{w} |^2}{1 +  \rho | \mathbf{h}^H \mathbf{q} |^2}\right)\right)\right)\right] + \nonumber
\end{array}
\end{equation}

\vspace{-3pt}

\begin{equation}
\begin{array}{l}
\hspace{-25pt} E \left[\log_2\left(1 + \exp\left(\ln\left(\frac{\rho | \mathbf{g}^H \mathbf{q} |^2}{1 +  \rho | \mathbf{g}^H \mathbf{w} |^2}\right)\right)\right)\right] \nonumber
\end{array}
\end{equation}

\vspace{-10pt}

\begin{small}
\begin{equation}
\begin{array}{l}
\vspace{35pt} \overset{(a)}{\ge} \log_2 \left[1 + \exp \Big ( E \left( \ln \left(\rho | \mathbf{h}^H \mathbf{w} |^2\right)\right) -  E \left( \ln \left(1 +  \rho | \mathbf{h}^H \mathbf{q} |^2\right)\right) \Big ) \right] \nonumber
\end{array}
\end{equation}
\end{small}

\vspace{-55pt}

\begin{small}
\begin{equation}
\begin{array}{l}
\vspace{18pt} + \; \log_2 \left[ 1 + \exp \Big ( E \left( \ln \left(\rho | \mathbf{g}^H \mathbf{q} |^2\right)\right) -  E \left( \ln \left(1 +  \rho | \mathbf{g}^H \mathbf{w} |^2\right)\right) \Big ) \right] \nonumber
\end{array}
\end{equation}
\end{small}

\begin{equation}
\hspace{-45pt} \overset{(b)}{\approx} \log_2 \left(1 + \frac{\rho \mathbf{w}^H \mathbf{R}_{\scriptscriptstyle{A}} \mathbf{w} }{\rho \mathbf{q}^H \mathbf{R}_{\scriptscriptstyle{A}} \mathbf{q} }\right) + \left(1 + \frac{\rho \mathbf{q}^H \mathbf{R}_{\scriptscriptstyle{B}} \mathbf{q} }{\rho \mathbf{w}^H \mathbf{R}_{\scriptscriptstyle{B}} \mathbf{w} }\right) \nonumber
\end{equation}

\begin{equation}
\hspace{-115pt} \overset{(c)}{\ge} \log_2 \left( \frac{\mathbf{w}^H \mathbf{R}_{\scriptscriptstyle{A}} \mathbf{w} }{\mathbf{w}^H \mathbf{R}_{\scriptscriptstyle{B}} \mathbf{w} } \frac{\mathbf{q}^H \mathbf{R}_{\scriptscriptstyle{B}} \mathbf{q} }{\mathbf{q}^H \mathbf{R}_{\scriptscriptstyle{A}} \mathbf{q} }\right). \nonumber
\end{equation}

Since $\log_2 (1 + r e^x )$ is convex in $x$ for $r > $0, we can obtain $(a)$ with Jensen's inequality. At high SNR, $(b)$ can be asymptotically approximated by first dropping `1 +' in the parentheses and applying Lemma 1. The tightness of $(b)$ has been shown in the asymptotic regime ($M \rightarrow \infty$) \cite{choi2013}. Moreover, the lower bound in $(c)$ is tight in high-correlated system with proper beamforming vectors. Interestingly, a recent work \cite{Qzhang2014} independently proved that $R_{\scriptstyle{\text{sum}}}$ can be well approximated by $(b)$ in massive MIMO system.

With $(c)$ at hand, we can transform the optimization problem into \vspace{-5pt}

\begin{equation} \label{eq:GEopt}
\underset{\|\mathbf{w}\|=1, \|\mathbf{q}\|=1}{ \text{max}} R_{\text{sum,lb}} \triangleq \log_2 \left( \frac{\mathbf{w}^H \mathbf{R}_{\scriptscriptstyle{A}} \mathbf{w} }{\mathbf{w}^H \mathbf{R}_{\scriptscriptstyle{B}} \mathbf{w} } \frac{\mathbf{q}^H \mathbf{R}_{\scriptscriptstyle{B}} \mathbf{q} }{\mathbf{q}^H \mathbf{R}_{\scriptscriptstyle{A}} \mathbf{q} }\right),
\end{equation}
for which the generalized eigenvector structure is the optimal solution \cite{wiesel2006}, as shown in eq. \eqref{eq:GE_precoder}. $\mathbf{w}$ corresponds to the dominant eigenvector of $\mathbf{R}^{-1}_{\scriptscriptstyle{B}}\mathbf{R}_{\scriptscriptstyle{A}}$ while $\mathbf{q}$ corresponds to the weakest one. The corresponding ergodic sum-rate satisfies $R_{\text{sum,lb}} = \log_2\left({\chi{\left(\mathbf{R}^{-1}_{\scriptscriptstyle{B}}\mathbf{R}_{\scriptscriptstyle{A}}\right)}}\right)$. Both $\mathbf{R}^{-1}_{\scriptscriptstyle{B}}\mathbf{R}_{\scriptscriptstyle{A}}$ and $\mathbf{R}^{-1}_{\scriptscriptstyle{A}}\mathbf{R}_{\scriptscriptstyle{B}}$ are positive definite, since $\mathbf{R}_{\scriptscriptstyle{A}}$,$\mathbf{R}_{\scriptscriptstyle{B}} \in \Phi_{PD}$ and $\left(\mathbf{R}^{-1}_{\scriptscriptstyle{B}}\mathbf{R}_{\scriptscriptstyle{A}}\right)^{-1} = \mathbf{R}^{-1}_{\scriptscriptstyle{A}}\mathbf{R}_{\scriptscriptstyle{B}}$. It is easy to find that $\chi{\left(\mathbf{R}^{-1}_{\scriptscriptstyle{A}}\mathbf{R}_{\scriptscriptstyle{B}}\right)} = \chi{\left(\mathbf{R}^{-1}_{\scriptscriptstyle{B}}\mathbf{R}_{\scriptscriptstyle{A}}\right)}$ and thereby we can obtain Theorem 1.

\section{PROOF OF PROPOSITION 1} \label{sec:proof_prop1}

We can lower bound the mutual information in \eqref{eq:amat_rate} applying \emph{Minkowski Determinant Theorem} \cite{marcus1992}
\vspace{-5pt}

\begin{eqnarray} \label{eq:mutual_infor}
I_{\scriptscriptstyle{A}} &=& \log_2  \det\left( \mathbf{I}_{2 \times 2} + \rho \, \mathbf{M} \right)\\
&\ge& \log_2 \left( 1 + \rho \, \det \left(\mathbf{M}\right)^{1/2} \right)^2 \\
&=& 2 \log_2 \left[ 1 + \rho \exp \left( \frac{1}{2} \ln \det \left(\mathbf{M}\right) \right) \right],
\end{eqnarray}
where
\vspace{-5pt}

\begin{small}
\begin{eqnarray} \label{eq:M_decomp}
\mathbf{M} &\triangleq& \widetilde{\mathbf{H}}^H \mathbf{K}^{-1} \widetilde{\mathbf{H}}\\
&=& \begin{bmatrix} \mathbf{W}^H \mathbf{h}_1, \mathbf{W}^H \mathbf{g}_1  \end{bmatrix}
\begin{bmatrix} \frac{|h_{21}|^{2}}{1 + |h_{21}|^{2}} & 0 \\ 0 & |h_{\scriptscriptstyle{31}}|^{2} \end{bmatrix}
\begin{bmatrix} \mathbf{h}^H_1 \mathbf{W} \\ \mathbf{g}^H_1 \mathbf{W} \end{bmatrix} \\
&=& \widetilde{\mathbf{G}}_{\scriptscriptstyle{2 \times 2}} \, \mathbf{\Lambda} \, \widetilde{\mathbf{G}}^H_{\scriptscriptstyle{2 \times 2}}.
\end{eqnarray}
\end{small}

By applying the convexity of $\log_2 (1 + r e^x ), r > 0$ and Jensen's inequality, the ergodic rate of user A per slot can be lower bounded as
\vspace{-5pt}

\begin{small}
\begin{eqnarray} \label{eq:amat_ergodic_rate}
R_{\scriptscriptstyle{A}}&\ge& \frac{2}{3} \, E \bigg\{\log_2\left[1 +\rho \exp \left( \frac{1}{2} \ln \det \left(\mathbf{M}\right) \right) \right]\bigg \}\\
&\ge& \frac{2}{3}  \log_2\left[1 +\rho \exp \left( \frac{1}{2} E \left[ \ln \det \left(\mathbf{M}\right) \right] \right) \right] \label{amat_rate_lb},
\end{eqnarray}
where $E \left[ \ln \det \left(\mathbf{M}\right) \right] = E \left[ \ln \det \left(\mathbf{\Lambda} \right) \right] +
E \left[ \ln \det \left(\widetilde{\mathbf{G}} \widetilde{\mathbf{G}}^H \right) \right]$. The first term can be further calculated with equations in \cite{gradshteyn1965}
\end{small}

\vspace{-5pt}

\begin{small}
\begin{eqnarray} \label{eq:amat_lambda}
\hspace{-10pt} E \left[ \ln \det \left(\mathbf{\Lambda}  \right) \right] &=& E \left[ \ln \left( \frac{|h_{\scriptscriptstyle{21}}|^{2}}{1 + |h_{\scriptscriptstyle{21}}|^{2}} \right) \right] + E \left[ \ln \left( |h_{\scriptscriptstyle{31}}|^{2}\right)\right] \\
&=& e \, \text{Ei}(-1) - 2\gamma \label{lambda_1},
\end{eqnarray}
\end{small}where (\ref{lambda_1}) is obtained by using the fact that $|h_{jm}|^2 \sim \text{Exp}(1)$. In general, it is nontrivial to evaluate the second term. A special case lies in i.i.d Rayleigh fading channel where $E \left[ \ln \det \left(\widetilde{\mathbf{G}} \widetilde{\mathbf{G}}^H \right) \right]$ can be exactly solved by invoking central Wishart distribution \cite{matthaiou2011}. For spatially correlated channel, we use Jensen's inequality to upper bound the second term as

\begin{equation}
\hspace{-70pt} E \left[ \ln \det \left(\tilde{\mathbf{G}} \tilde{\mathbf{G}}^H \right) \right] \le \ln E \left[ \det \left(\tilde{\mathbf{G}} \tilde{\mathbf{G}}^H \right) \right] \nonumber
\end{equation}

\vspace{-12pt}

\begin{spacing}{1.5}
\begin{equation}
\begin{array}{l}
\hspace{23pt} \overset{(a)}{=} \ln [ E (\mathbf{h}^H_1 \mathbf{W} \mathbf{W}^H \mathbf{h}_1 \mathbf{g}^H_1 \mathbf{W} \mathbf{W}^H \mathbf{g}_1  - \nonumber
\\ \hspace{60pt} \mathbf{h}^H_1 \mathbf{W} \mathbf{W}^H \mathbf{g}_1 \mathbf{g}^H_1 \mathbf{W} \mathbf{W}^H \mathbf{h}_1 ) ] \nonumber
\end{array}
\end{equation}
\end{spacing}

\vspace{-25pt}

\begin{spacing}{1.5}
\begin{equation}
\begin{array}{l}
\hspace{50pt} \overset{(b)}{=} \ln [ E \left(\mathbf{h}^H_1 \mathbf{W} \mathbf{W}^H \mathbf{h}_1\right) E\left( \mathbf{g}^H_1 \mathbf{W} \mathbf{W}^H \mathbf{g}_1\right)  - \nonumber
\\ \hspace{72pt} E \left( \mathbf{h}^H_1 \mathbf{W} \mathbf{W}^H \mathbf{g}_1 \mathbf{g}^H_1 \mathbf{W} \mathbf{W}^H \mathbf{h}_1 \right) ] \nonumber
\end{array}
\end{equation}
\end{spacing}

\vspace{-12pt}

\begin{equation}
\begin{array}{l}
\hspace{-72pt} \overset{(c)}{=} \ln (\Theta_{\scriptscriptstyle{\text{A}}}), \label{eq:amat_wishart}
\end{array}
\end{equation}
where $\Theta_{\scriptscriptstyle{\text{A}}}$ is defined in \eqref{eq:theta}. Eq. $(a)$ is obtained with $\det (\mathbf{A}\mathbf{B}) = \det (\mathbf{A})\det(\mathbf{B})$ for equal-size square matrices $\mathbf{A},\mathbf{B}$. Eq. $(b)$ is because $\mathbf{h}_1$ and $\mathbf{g}_1$ are independent Gaussian random vectors. Noting that $\text{Tr}(\mathbf{A}\mathbf{B}) = \text{Tr}(\mathbf{B}\mathbf{A})$ and $E\left[\text{Tr}\left(\mathbf{C}\right)\right] = \text{Tr}\left[E\left(\mathbf{C}\right)\right]$, eq. $(c)$ can be easily calculated.

Finally, substituting (\ref{lambda_1}) and \eqref{eq:amat_wishart} into (\ref{amat_rate_lb}) renders an analytical approximation of the ergodic rate of user A and hence completes the proof.

\section{PROOF OF PROPOSITION 3} \label{sec:Proof_props2}

Define $\mathbf{M} \triangleq \widetilde{\mathbf{H}}_1^H \mathbf{K}^{-1} \widetilde{\mathbf{H}}_1 \mathbf{P}^2_{\scriptscriptstyle{A}}$ and with simple manipulations, we have

\vspace{-5pt}

\begin{small}
\begin{equation}
\mathbf{M} = [\mathbf{W}^H \mathbf{h}_1, \mathbf{W}^H \mathbf{g}_1]
\begin{bmatrix} \frac{1}{k_1} + \frac{P_{\scriptscriptstyle{5}} |h_{21}|^{2}}{k_2} & 0 \\ 0 & \frac{P_{\scriptscriptstyle{8}} |h_{31}|^{2}}{k_3}  \end{bmatrix}
\begin{bmatrix} \mathbf{h}^H_1 \mathbf{W} \\ \mathbf{g}^H_1 \mathbf{W} \end{bmatrix} \mathbf{P}^2_{\scriptscriptstyle{A}}
\end{equation}
\end{small}

\vspace{-20pt}

\begin{small}
\begin{equation} \label{prop2_M}
\hspace{-143pt} = \tilde{\mathbf{G}}_{\scriptscriptstyle{2 \times 2}} \, \mathbf{\Gamma} \, \tilde{\mathbf{G}}^H_{\scriptscriptstyle{2 \times 2}} \mathbf{P}^2_{\scriptscriptstyle{A}}.
\end{equation}
\end{small}

Rewrite \eqref{eq:smat_amat} as

\vspace{-22pt}

\begin{spacing}{1.5}
\begin{eqnarray} \label{prop2_proof2}
R_{\mathbf{s}_{\scriptscriptstyle{A}}} &=& E \left[ \log_2  \det\left( \mathbf{I}_{2\times2} + \mathbf{M}  \right)\right] \label{prop2_eq1} \\
&=& E \left[ \log_2  \det\left(1 + \text{Tr}(\mathbf{M}) + \det(\mathbf{M})  \right)\right] \label{prop2_eq2} \\ \label{prop2_eq3}
&\le& \log_2  \det [1 + E \left(\text{Tr}\left(\mathbf{M}\right)\right) + E\left(\det\left(\mathbf{M} \right)\right)].
\end{eqnarray}
\end{spacing}

Eq. \eqref{prop2_eq1} is obtained with $\det(\mathbf{I} + \mathbf{A}\mathbf{B}) = \det(\mathbf{I} + \mathbf{B}\mathbf{A})$ while \eqref{prop2_eq2} makes use of the Cayley-Hamilton theorem. Then, we upper bound \eqref{prop2_eq2} by \eqref{prop2_eq3} using Jensen's inequality. With the help of \eqref{prop2_M} and $E[\text{Tr}(\cdot)] = \text{Tr}[E(\cdot)]$, the first term in \eqref{prop2_eq3} can be characterized as
\vspace{-5pt}

\begin{small}
\begin{equation} \label{prop2_proof3}
E \left[\text{Tr}\left(\mathbf{M}\right)\right] = \text{Tr}\left[\left( E \left( \Gamma_{11} \right) \mathbf{W}^H \mathbf{R}_{\scriptscriptstyle{A}} \mathbf{W} + E \left( \Gamma_{22} \right) \mathbf{W}^H \mathbf{R}_{\scriptscriptstyle{B}} \mathbf{W} \right) \mathbf{P}^2_{\scriptscriptstyle{A}} \right]
\end{equation}
\end{small}

\vspace{-20pt}

\begin{small}
\begin{equation} \label{prop2_para1}
\hspace{35pt} \approx \delta_{\scriptscriptstyle{A1}} \left(\tau_{\scriptscriptstyle{A1}}P_{\scriptscriptstyle{1}} + \tau_{\scriptscriptstyle{A2}}P_{\scriptscriptstyle{2}}\right) + \delta_{\scriptscriptstyle{A2}} \left(\lambda_{\scriptscriptstyle{B1}}P_{\scriptscriptstyle{1}} + \lambda_{\scriptscriptstyle{B2}}P_{\scriptscriptstyle{2}} \right),
\end{equation}
\end{small}where $\lambda_{\scriptscriptstyle{B1}} = \mathbf{w}^H_1 \mathbf{R}_{\scriptscriptstyle{B}} \mathbf{w}_1, \lambda_{\scriptscriptstyle{B2}} = \mathbf{w}^H_2 \mathbf{R}_{\scriptscriptstyle{B}} \mathbf{w}_2, \tau_{\scriptscriptstyle{A1}} = \mathbf{w}^H_1 \mathbf{R}_{\scriptscriptstyle{A}} \mathbf{w}_1$, $\tau_{\scriptscriptstyle{A2}} = \mathbf{w}^H_2 \mathbf{R}_{\scriptscriptstyle{A}} \mathbf{w}_2$ and

\begin{small}
\begin{equation} \label{prop2_delta1}
\hspace{-5pt} E \left( \Gamma_{11} \right) = E \left(\frac{1}{k_1} + \frac{P_{\scriptscriptstyle{5}} |h_{\scriptscriptstyle{21}}|^{2}}{k_2}\right),
E \left( \Gamma_{22} \right) = E \left(\frac{P_{\scriptscriptstyle{8}} |h_{\scriptscriptstyle{31}}|^{2} }{k_3} \right).
\end{equation}
\end{small}

The terms on the right hand side of \eqref{prop2_delta1} can be further evaluated as follows:

\vspace{-2pt}

\begin{small}
\begin{eqnarray} \label{prop2_delta2}
\hspace{-20pt} E \left(\frac{1}{k_1}\right) &=&  E \left( \frac{1}{1 + |\mathbf{h}^H_1 \mathbf{Q} \mathbf{P}_{\scriptscriptstyle{B}}|^2}\right) \nonumber \\
&\overset{a}{\ge}& \frac{1}{1 +  E \left( |\mathbf{h}^H_1 \mathbf{Q} \mathbf{P}_{\scriptscriptstyle{B}}|^2\right) } = \frac{1}{1+\lambda_{\scriptscriptstyle{A1}}P_{\scriptscriptstyle{3}} + \lambda_{\scriptscriptstyle{A2}}P_{\scriptscriptstyle{4}}} \label{delta21}
\end{eqnarray}
\end{small}

\vspace{-15pt}

\begin{small}
\begin{eqnarray}
\hspace{-20pt} E \left( \frac{P_{\scriptscriptstyle{5}} |h_{\scriptscriptstyle{21}}|^{2}}{k_2}\right)&=&E \left( \frac{\mathbf{h}^H_{w,2} \mathbf{A} \mathbf{h}_{w,2}}{1+\mathbf{h}^H_{w,2} \mathbf{B} \mathbf{h}_{w,2}}\right) \nonumber \\
&\overset{b}{\approx}& \frac{\text{Tr}(\mathbf{A})}{1+\text{Tr}(\mathbf{B})} = \frac{P_{\scriptscriptstyle{5}}} {1+P_{\scriptscriptstyle{5}}+ \tau_{\scriptscriptstyle{A3}}P_{\scriptscriptstyle{6}}+ \lambda_{\scriptscriptstyle{A3}}P_{\scriptscriptstyle{7}} }\label{delta22}
\end{eqnarray}
\end{small}where $\lambda_{\scriptscriptstyle{A1}} = \mathbf{q}^H_1 \mathbf{R}_{\scriptscriptstyle{A}} \mathbf{q}_1, \, \lambda_{\scriptscriptstyle{A2}} = \mathbf{q}^H_2 \mathbf{R}_{\scriptscriptstyle{A}} \mathbf{q}_2, \, \tau_{\scriptscriptstyle{A3}} = \mathbf{w}^H_3 \mathbf{R}_{\scriptscriptstyle{A}} \mathbf{w}_3$, $\lambda_{\scriptscriptstyle{A3}} = \mathbf{q}^H_3 \mathbf{R}_{\scriptscriptstyle{A}} \mathbf{q}_3$. Inequality $(a)$ comes from the fact that $\frac{1}{x}$ is convex in $x$ for $x>0$. Note that \eqref{delta21} can be exactly calculated as an exponential integral function of $\lambda_{\scriptscriptstyle{A1}} P_{\scriptscriptstyle{3}}, \, \lambda_{\scriptscriptstyle{A2}}P_{\scriptscriptstyle{4}}$. Nevertheless, such implicit characterization restrains insightful analysis of the power allocation strategy (for instance, how the power assigned to signal of user B interferes user A).

In \eqref{delta22}, $\mathbf{A} = P_{\scriptscriptstyle{5}} \mathbf{R}^{1/2}_{\scriptscriptstyle{A}} \mathbf{x}_1 \mathbf{x}_1^H \mathbf{R}^{1/2}_{\scriptscriptstyle{A}}$ where $\mathbf{x}_1 = [1,\,0]^T$ and $\mathbf{B} = \mathbf{R}^{1/2}_{\scriptscriptstyle{A}} (P_{\scriptscriptstyle{5}} \mathbf{x}_1 \mathbf{x}_1^H + P_{\scriptscriptstyle{6}} \mathbf{w}_1 \mathbf{w}_1^H + P_{\scriptscriptstyle{7}} \mathbf{q}_1 \mathbf{q}_1^H) \mathbf{R}^{1/2}_{\scriptscriptstyle{A}}$. $(b)$ is based on the first order approximation in \eqref{taylor_exp1}. The second (and higher) order approximation would be more accurate, however, rendering the problem too complicated to implement optimization techniques\footnote{For instance, it is difficult to compute the first/second order derivatives of the objective function which are necessary for various non-linear programming methods.}. Similarly, we can approximate $E \left( \Gamma_{22} \right)$ as

\vspace{-5pt}

\begin{eqnarray} \label{prop2_delta3}
E \left(\frac{P_{\scriptscriptstyle{8}} |h_{\scriptscriptstyle{31}}|^{2} }{k_3} \right) &=& E \left( \frac{\mathbf{h}^H_{w,3} \mathbf{C} \mathbf{h}_{w,3}}{1+\mathbf{h}^H_{w,3} \mathbf{D} \mathbf{h}_{w,3}}\right) \approx \frac{\text{Tr}(\mathbf{C})}{1+\text{Tr}(\mathbf{D})} \nonumber \\
&=& \frac{P_{\scriptscriptstyle{8}}} {1+\tau_{\scriptscriptstyle{A3}} P_{\scriptscriptstyle{9}} + \lambda_{\scriptscriptstyle{A3}}P_{\scriptscriptstyle{10}} }.
\end{eqnarray}

$\mathbf{D} = \mathbf{R}^{1/2}_{\scriptscriptstyle{A}} (P_{\scriptscriptstyle{9}} \mathbf{w}_1 \mathbf{w}_1^H + P_{\scriptscriptstyle{10}} \mathbf{q}_1 \mathbf{q}_1^H) \mathbf{R}^{1/2}_{\scriptscriptstyle{A}}$ and $\mathbf{C} = P_{\scriptscriptstyle{8}} \mathbf{R}^{1/2}_{\scriptscriptstyle{A}} \mathbf{x}_1 \mathbf{x}_1^H \mathbf{R}^{1/2}_{\scriptscriptstyle{A}}$. The second term in \eqref{prop2_eq3} can be given by

\vspace{-5pt}

\begin{small}
\begin{equation} \label{prop2_proof4}
E \left[\det\left(\mathbf{M}\right)\right] = E \left[\det\left(\mathbf{\Gamma}\right)\right] \cdot E \left[\det\left(\tilde{\mathbf{G}} \,\tilde{\mathbf{G}}^H \right)\right] \cdot E \left[\det\left(\mathbf{P}^2_{\scriptscriptstyle{A}} \right)\right]
\end{equation}
\end{small}

\vspace{-20pt}

\begin{equation}\label{prop2_para2}
\hspace{-80pt} \approx \delta_{\scriptscriptstyle{A1}} \delta_{\scriptscriptstyle{A2}} \Theta_{\scriptscriptstyle{\text{A}}} P_{\scriptscriptstyle{1}}P_{\scriptscriptstyle{2}},
\end{equation}where calculation of $E \left[\det\left(\widetilde{\mathbf{G}} \,\widetilde{\mathbf{G}}^H \right)\right]$ follows $(a),(b),(c)$ of eq. \eqref{eq:amat_wishart}. Substituting \eqref{prop2_delta2} $\sim$ \eqref{prop2_delta3} into \eqref{prop2_proof3} and \eqref{prop2_proof4}, we can obtain \eqref{prop2_para1} and \eqref{prop2_para2}. Combining \eqref{prop2_para1} and \eqref{prop2_para2} with \eqref{prop2_eq3} establishes \eqref{eq:smat_rate1}.

In order to compute $R^p_{\mathbf{s}_{\scriptscriptstyle{A}}}$, we can reexpress \eqref{eq:smat_sbf} as

\begin{small}
\begin{eqnarray} \label{eq:smat_sbf_cal}
\hspace{-50pt} R^p_{\mathbf{s}_{\scriptscriptstyle{A}}} &=& E \left[ \log_2 \left(1+\frac{P_{\scriptscriptstyle{6}} |\mathbf{h}^H_2 \mathbf{w}_3|^2}{1+P_{\scriptscriptstyle{5}}|h_{\scriptscriptstyle{21}}|^2 +P_{\scriptscriptstyle{7}} |\mathbf{h}^H_2 \mathbf{q}_3|^2}  \right) \right] + \nonumber \\
&& E \left[ \log_2 \left(1+\frac{P_{\scriptscriptstyle{9}} |\mathbf{h}^H_3 \mathbf{w}_3|^2}{1+P_{\scriptscriptstyle{10}} |\mathbf{h}^H_3 \mathbf{q}_3|^2}  \right) \right] \label{sbf_cal1} \\
&\le& \log_2 \left[1+ E \left( \frac{P_{\scriptscriptstyle{6}} |\mathbf{h}^H_2 \mathbf{w}_3|^2}{1+P_{\scriptscriptstyle{5}}|h_{\scriptscriptstyle{21}}|^2 + P_{\scriptscriptstyle{7}} |\mathbf{h}^H_2 \mathbf{q}_3|^2}  \right) \right] + \nonumber \\
&& \log_2 \left[1+  E \left(\frac{P_{\scriptscriptstyle{9}} |\mathbf{h}^H_3 \mathbf{w}_3|^2}{1+P_{\scriptscriptstyle{10}} |\mathbf{h}^H_3 \mathbf{q}_3|^2} \right) \right] \label{sbf_cal2}
\end{eqnarray}
\end{small}

\vspace{-20pt}

\begin{small}
\begin{equation}
\approx \log_2 \left(1 + \frac{\tau_{\scriptscriptstyle{A3}}P_{\scriptscriptstyle{6}}}{1+P_{\scriptscriptstyle{5}}+\lambda_{\scriptscriptstyle{A3}}P_{\scriptscriptstyle{7}}} \right) + \log_2 \left(1 + \frac{\tau_{\scriptscriptstyle{A3}}P_{\scriptscriptstyle{9}}} {1+\lambda_{\scriptscriptstyle{A3}}P_{\scriptscriptstyle{10}}} \right).
\end{equation}
\end{small}

An analytical expression of \eqref{sbf_cal1} was obtained for the case $M = 2$ in \cite{raghavan2011}, while a lower bound for $M > 2$ case is derived in section \ref{sec:SBF}. We here use Jensen's inequality and \eqref{taylor_exp1} in Lemma 2 to estimate \eqref{sbf_cal1}, leading to an approximation \eqref{sbf_cal2} as well as \eqref{eq:smat_rate1}.

\bibliographystyle{IEEEtran}
\bibliography{reference}

\end{document}